\documentclass[aps,pra,showpacs,twoside,10pt,floatfix,nofootinbib,longbibliography,twocolumn]{revtex4-1}
\usepackage[colorlinks=true, citecolor=red, urlcolor=blue ]{hyperref}
\usepackage{epsfig,newlfont,amssymb,amsfonts,amsmath,bm,subfigure,palatino,mathtools,amsthm,braket,soul,enumitem,color,times,comment,geometry,float,array}
\usepackage[table]{xcolor}
\usepackage[normalem]{ulem}
\begin{document}

\title{Entanglement and fidelity across quantum phase transitions in locally perturbed topological codes with open boundaries}
\author{Harikrishnan K J and Amit Kumar Pal}
\affiliation{Department of Physics, Indian Institute of Technology Palakkad, Palakkad 678 623, India}
\date{\today}

\begin{abstract}
We investigate the topological-to-non-topological quantum phase transitions (QPTs) occurring in the Kitaev code under local perturbations in the form of local magnetic field and spin-spin interactions of the Ising-type using fidelity susceptibility (FS) and entanglement as the probes. We assume the code to be embedded on the surface of a wide cylinder of height $M$ and circumference $D$ with $M\ll D$. We demonstrate a power-law divergence of FS across the QPT, and determine the quantum critical points (QCPs) via a finite-size scaling analysis. We verify these results by mapping the  perturbed Kitaev code to the 2D Ising model with nearest- and next-nearest-neighbor interactions, and computing the single-site magnetization  as order parameter using quantum Monte-Carlo technique. We also demonstrate a finite size odd-even dichotomy in the occurrence of the QPT in the Kitaev ladder with respect to the odd and even values of $D$, when the system is perturbed with only Ising interaction. Our results also indicate a higher robustness of the topological phase of the Kitaev code against local perturbations if the boundary is made open along one direction. We further consider a local entanglement witness operator designed specifically to capture a lower bound to the localizable  entanglement  on the vertical non-trivial loop of the code. We show that the first derivative of the expectation value of the witness operator exhibits a logarithmic divergence across the QPT, and perform the finite-size scaling analysis. We demonstrate similar behaviour of the expectation value of the appropriately constructed witness operator also in the case of locally perturbed color code with open boundaries.     
\end{abstract}

\maketitle

\section{Introduction}
\label{intro}

Investigating novel phases and corresponding quantum phase transitions (QPTs)~\cite{Sachdev2011} in quantum many-body systems with correlations arising out of quantum information science has arguably been one of the most prominent research topics in the last two decades. On one hand, the Landau paradigm~\cite{Wilson1974,*Landau1999} of phases characterized by local order parameters, and the corresponding QPTs associated with spontaneous symmetry breaking~\cite{Sachdev2011} have been studied extensively in terms of fidelity susceptibility (FS)~\cite{You2007,*Jacobson_2009,*Gu_2010} and entanglement~\cite{amico2008,*Latorre2009,*dechiara2018}, including \emph{local} entanglement over a subsystem of size far less than that of the entire system~\cite{osborne2002,*osterloh2002,amico2008,*Latorre2009,*dechiara2018} as well as  \emph{global} entanglement that considers the state of the entire system~\cite{Deger_2019,amico2008,*Latorre2009,*dechiara2018}.  On the other hand, there exist topological phases (TPs) outside the Landau paradigm -- relevant particularly in studying fractional quantum Hall effect~\cite{wen1990,*wen1995,*Stormer1999}, quantum spin liquids~\cite{Isakov2011,*White2011},
topological quantum codes~\cite{kitaev2001,*kitaev2003,*kitaev2006,bombin2006,*bombin2007}, and quantum walks~\cite{Kitagawa_2010,*Grudka_2023} -- which can not be probed with local order parameters, and has been explored using fidelity~\cite{Garnerone_2009,*Panahiyan_2020} and topological entanglement entropy~\cite{kitaev2006a,*levin2006}.

\begin{figure*}
    \includegraphics[width=0.9\linewidth]{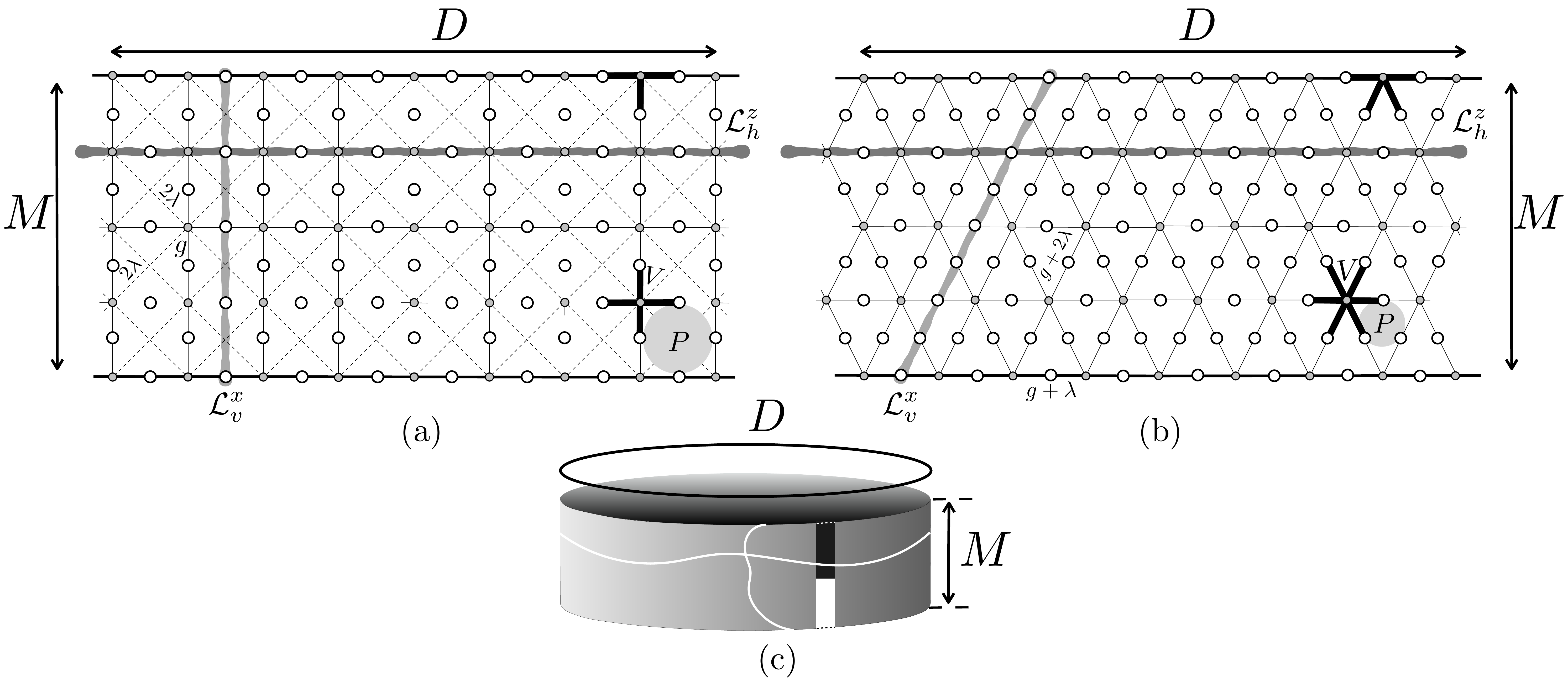}
    \caption{Kitaev code defined on the (a) square and the (b) triangular lattices with open (periodic) boundary condition along the vertical (horizontal) directions, where qubits (represented by white circles) sit on the edges (represented by thin lines) of the lattice. Each of the $N_P$ plaquette operators, indexed by $P$ (denoted by shaded regions), involves respectively $4$ and $3$ qubits for the square and the triangular lattices, and each of the $N_V$ vertex operators, $V$ (shown by solid gray circles and thick black lines), involves $4$ and $6$ qubits respectively in the bulk, and $3$ and $4$ qubits respectively at the boundary.  The non-trivial loops $L^{x,z}_{h,v}$ corresponding to the logical operators $\mathcal{L}^{x,z}_{h,v}$, are shown with thick lines, where $h$ and $v$ in the subscript denote the horizontal and vertical directions. We consider the 2D lattice  in the shape of a cylinder of circumference $D$ and height $M$ with $D\gg M$ (c), where only two non-trivial loops -- one $x$-type in the vertical direction, and the other $z$-type in the horizontal direction -- exist. The interaction strengths corresponding to different NN and NNN (shown by dashed thin lines) qubit-pairs in the mapped transverse-field Ising model $\tilde{H}$ are given by $g$ and $2\lambda$ respectively for the square lattice, where only NN qubit-pairs interact via a strength $g+2\lambda$ in the bulk, and $g+\lambda$ in the boundary of the triangular lattice.}
    \label{fig:kitaev_toric_code}
\end{figure*}

The \emph{topological-to-non-topological}  QPTs occurring in the lattice models referred to as the \emph{topological quantum codes} (TQCs),  including the Kitaev code~\cite{kitaev2001,*kitaev2003,*kitaev2006} and the color code~\cite{bombin2006,*bombin2007}, in the presence of external perturbations has attracted a lot of attention due to the importance of these models in topological quantum computation~\cite{nayak2008}. Such a QPT results in the disappearance of the  TP of the lattice models due to local perturbations in the form of local magnetic fields, or spin-spin interactions
~\cite{Trebst2007,*Vidal2009_1,*Dusuel2011,*Wu2012,*Tsomokos2011,*Zarei2019a,Karimipour2013,Zarei2015,Jahromi2013,*Jahromi2013a,Jamadagni2018}, thereby quantifying the \emph{robustness} of the TPs.  So far, such QPTs have mostly been probed using ground state energy per site~\cite{Vidal2009_1,Dusuel2011,Jahromi2013}, expectation value of Wilson loops~\cite{Karimipour2013}, energy gaps~\cite{Vidal2009_1,Dusuel2011,Trebst2007,Jahromi2013,Jamadagni2018},  and topological entanglement entropy~\cite{Jamadagni2018}. Probing QPTs occurring in quantum many-body systems using FS exploits the fact that the ground state (GS) of a quantum many-body Hamiltonian undergoes a drastic change at the quantum critical point (QCP), and therefore exhibits a sharp drop in the fidelity between the ground states of the model at two different, and yet very close instances of the system parameter bringing about the QPT~\cite{You2007,*Jacobson_2009,*Gu_2010}. While FS as a probe appears to be applicable for all QPTs, recent observation of the shifting of the peak in  FS from the QCP corresponding to the Berezinskii-Kosterlitz-Thouless transition~\cite{Cincio_2019} poses the question as to whether the QPTs occurring in the topological codes can be faithfully captured by the FS. However, study of such QPTs using FS is challenging due to the difficulty in fully accessing the GSs of the perturbed TQCs.

The GSs in the TP (TPGSs) of the Kitaev code~\cite{kitaev2001,*kitaev2003,*kitaev2006} and the color code~\cite{bombin2006,*bombin2007} are genuinely multiparty entangled (see ~\cite{horodecki2009,Hein2005} and the references therein) \emph{stabilizer} states~\cite{nielsen2010}. Due to the connection of such states with graph states~\cite{Hein2005} via local Clifford operations~\cite{van-den-nest2004,*Lang2012}, full information regarding the entanglement of this state is accessible~\cite{Hein2005,HK2023}. However, in the presence of perturbation in the form of local magnetic field or spin-spin interactions, it is not so. In this paper, we focus on the local entanglement over a subset of qubits. In the case of stabilizer states, the local entanglement over a chosen subset of qubits has been shown to be  efficiently quantified using only a measurement-based approach~\cite{verstraete2004,*verstraete2004a,*popp2005,Hein2005,amaro2018,*Amaro2020a,HK2023,HK2022}, where strategic local Pauli measurements on all qubits outside the subset results in non-zero average entanglement -- bipartite or multipartite -- over the qubits in the selected subset. This average entanglement maximized over all possible local Pauli measurements on all qubits outside the subset is referred to as the \emph{localizable entanglement} (LE)~\cite{verstraete2004,*verstraete2004a,*popp2005}. While this motivates one to compute LE for probing the QPTs in locally perturbed topological codes, the optimal Pauli measurement setup for LE over arbitrary subsystems can be determined only in the case of vanishing perturbations~\cite{amaro2018,*Amaro2020a,HK2023}. When perturbation is present, similar to the FS, one needs to access the full state for determining LE over a chosen subsystem, and the problem becomes exponentially difficult for large system sizes\footnote{In fact, it is been shown for correlations such as quantum discord~\cite{Ollivier_2001,*Henderson_2001} which uses similar optimization as in the case of LE that the problem is NP-hard~\cite{Huang2014}.}.

We take up these challenges by considering the Kitaev code, defined on the square and the triangular lattices, in the presence of local perturbations having the form of single-qubit magnetic field and spin-spin interactions of the Ising type. The usually studied cases~\cite{Trebst2007,*Vidal2009_1,*Dusuel2011,*Wu2012,*Tsomokos2011,*Zarei2019a,Karimipour2013,Zarei2015,Jahromi2013,*Jahromi2013a} correspond to the full 2D lattices of dimension $M\times D$ with $M=D$, $M$ and $D$ being the sizes along the vertical and the horizontal directions, and periodic boundary condition (PBC) is assumed along both the horizontal and the vertical directions. In contrast,  we consider the code to be defined on lattices with a constant $M$ such that $M\ll D$, where a periodic boundary conditions (PBC) is assumed along the horizontal direction, and an open boundary condition (OBC) along the vertical. Therefore, the code can be viewed to be embedded on a \emph{wide} cylinder with length (or, height), $M$, and circumference, $D$ (see Fig.~\ref{fig:kitaev_toric_code}). We study the quantum critical behaviour of the perturbed quantum code on this lattice in the limit $D\rightarrow\infty$. This is in the same vein as the cases  of the \emph{quasi one-dimensional} (1D) quantum spin models, where the size of the system along, say, the vertical direction is much smaller than the size along the horizontal direction, and the thermodynamic limit is approached when the size in the horizontal (vertical) direction is increased (kept fixed)~\cite{Mila1998,*Tandon1999,*Ercolessi2003,*Batchelor2007,*Ivanov2009}. Our investigation is inspired from the quantitative, and at times, qualitative differences between the critical phenomena observed in the quasi-one dimensional (1D) models, and the same in the truly 1D, or full two-dimensional (2D) systems under similar perturbation~\cite{Mila1998,*Tandon1999,*Ercolessi2003,*Batchelor2007,*Ivanov2009}. Further, detection of the quantum critical points by entanglement measures in the quasi-1D models ~\cite{Tribedi2009} boosts our motivation of investigating the possibility of the same in topological quantum codes defined on quasi 1D lattices.

We determine the ground state of the perturbed Kitaev code using exact diagonalization (ED), and calculate the FS with respect to the field and the Ising perturbation strengths. Across the QCPs on the field and the Ising interaction axis, FS exhibits a power-law divergence in the limit $D\rightarrow\infty$. In order to overcome the limitation in the size of the system due to the use of ED and to verify QCPs determined using the FS, we map the perturbed Kitaev code with OBC along the vertical direction on to a 2D Ising model with nearest-neighbor (NN) and next-nearest-neighbor (NNN) interactions. We probe the QPT occurring in the perturbed Kitaev code using the single-site magnetization  of the 2D Ising model, computed using the continuous time quantum Monte-Carlo technique with cluster update~\cite{Rieger1999}. The results support the QCPs obtained using the FS. We perform similar investigation for the perturbed Kitaev code embedded on a cylinder with a length far greater than its circumference also, and determine the QCPs. Our results indicate  that the topological phase of the Kitaev code with OBC along the vertical direction is more robust against local perturbations in the form of the local field and the Ising interactions, as compared to the same in the case of the Kitaev code embedded on a torus with PBC along both directions. In the case of the Ising perturbation on square lattice, we further point out a \emph{finite size odd-even dichotomy} in the approach of the system towards the limit $D\rightarrow\infty$, manifested in how the finite-size system approaches the QPT when $D$ is gradually increased.

Next, we focus on the LE over the vertical non-trivial loop corresponding to the Kitaev code, and note that a lower bound of this LE  can be calculated via constructing an appropriate \emph{local} entanglement witness operator~\cite{Alba2010, Amaro2020} (cf.~\cite{guhne2009}) on the selected subset of qubits~\cite{amaro2018,*Amaro2020a,HK2022,HK2023}, in the same vein as an appropriately constructed witness operator provides a lower bound of a properly chosen entanglement measure on a quantum state~\cite{Brandao2005,*Brandao2006,*Eisert2007}. The construction of the witness operator depends on the optimal Pauli measurement setup on the qubits outside the subset~\cite{amaro2018,*Amaro2020a,HK2022,HK2023} for localizing maximum entanglement on the loop. Using ED, it has been shown~\cite{HK2022} in the case of TQCs of moderately small sizes perturbed with local magnetic field that the witness-based lower bound of LE has the potential to be a marker for the field-induced QPT. Nevertheless, due to obvious numerical limitations in accessing large systems, a full scaling analysis of LE across the QPTs of a  locally perturbed TQC remains a difficult problem.

To address this, we consider the local entanglement witness operator constructed specifically to capture a lower bound of the LE -- both bipartite and genuine multipartite -- on the subset of qubits forming the vertical non-trivial loop in the unperturbed Kitaev code. We determine the expectation value of the witness operator in the GS of the perturbed Kitaev code by exploiting its mapping to the 2D Ising model. We demonstrate that the first derivative of the expectation value of the local witness operator w.r.t. the strength of the local perturbation exhibits a logarithmic divergence across the QCP, and perform the corresponding finite-size scaling analysis. We further show persistence of such behaviour for the witness operator by extending our investigation to the color code perturbed locally by a parallel magnetic field~\cite{Jahromi2013} and embedded on a wide cylinder, where we use mapping of the model on to the 2D Baxter-Wu model with a transverse field~\cite{Capponi2014} and additional Ising interactions along the boundary for computing the expectation value of the constructed local witness operator.

The rest of the paper is organized as follows. In Sec.~\ref{sec:Kitaev}, we discuss the salient features of the Kitaev code embedded on the surface of a wide cylinder (Sec.~\ref{subsec:toric_code}). Determination of the QCPs of the locally perturbed Kitaev code using FS, and its verification with local magnetization of the mapped 2D Ising model is discussed in Sec.~\ref{subsec:kitaev_qpt_magnetization}.  Sec.~\ref{subsec:kitaev_entanglement} deals with the behaviour of the expectation value of the specially constructed local entanglement witness operator capturing a lower bound of LE over the vertical non-trivial loop of the Kitaev code. The persistence of similar behaviour by the local witness operator in the locally perturbed color code embedded on a wide cylinder is demonstrated in Sec.~\ref{sec:color}. Sec.~\ref{sec:conclusion} contains the concluding remarks and outlook.

\begin{table*}
    \centering
    \begin{tabular}{|wl{0.82\textwidth}|}
    \hline
    \cellcolor{blue!15} \textbf{Square lattice} \\  
    \begin{tabular}{wl{0.225\textwidth}|wl{0.15\textwidth}|wl{0.14\textwidth}|wl{0.27\textwidth}}
    \hline
       \textbf{Boundary Condition} & \textbf{Lattice geometry} & $g_c^0$ & $\lambda_c^0$ \\
    \hline 
      PBC along $h$ and $v$ & $D = M$ & $0.328$~\cite{Trebst2007,Vidal2009_1,Wu2012} & $0.166$~\cite{Karimipour2013,Zarei2015} \\
    \hline 
      PBC along $h$, OBC along $v$ & $M\ll D$ & $0.541$ & $0.494^*$ \\
    \hline 
     PBC along $h$, OBC along $v$ & $M\gg D$ & $0.368$ & $0.174$   
    \end{tabular} \\
    \hline 
    \cellcolor{blue!15} \textbf{Triangular lattice} \\
    \begin{tabular}{wl{0.225\textwidth}|wl{0.15\textwidth}|wl{0.14\textwidth}|wl{0.27\textwidth}}
    \hline 
       \textbf{Boundary Condition} & \textbf{Lattice geometry} & $g_c^0$ & $\lambda_c^0$ \\
    \hline 
       PBC along $h$ and $v$ & $D = M$ & $0.209$~\cite{Blote2002} & $0.104$~\cite{Zarei2015} \\
    \hline 
     PBC along $h$, OBC along $v$ & $M\ll D$ & $0.400$ & $0.276$ \\
    \hline  
      PBC along $h$, OBC along $v$ & $M\gg D$ & $0.231$ & $0.113$    
    \end{tabular}\\    
    \hline 
    \end{tabular}
    \caption{Values of $g_c^0$ and $\lambda_c^0$ corresponding to Kitaev code on the square and the triangular lattices of different geometries with different boundary conditions. The values corresponding to $M\ll D$ and $M\gg D$ limits are obtained using magnetization as the order parameter computed via the QMC technique. For $M\ll D$ limit, we set $M=2$ and $D\in[39,75]$ while for $M\gg D$ limit, we use $D=3$ and $M\in[15,51]$. The value of $\lambda_c^0$ marked with an asterisk, corresponding to the PBC (OBC) along $h$ ($v$) in the case of the square lattice is obtained considering only odd values of $D$ (See Appendix \ref{app:odd_even_dichotomy} for an explanation on the underlying subtleties and a discussion on a finite-size odd-even dichotomy). All quantities are dimensionless.}
    \label{tab:qpt}
\end{table*}

\section{Locally perturbed Kitaev codes}
\label{sec:Kitaev}

In this section, we focus on the locally perturbed Kitaev code embedded on a lattice with open boundary condition along one side, and study the phase transition induced by the perturbation.  

\subsection{Toric code under open boundary condition}
\label{subsec:toric_code}

\begin{figure*}
    \centering
    \includegraphics[width=\linewidth]{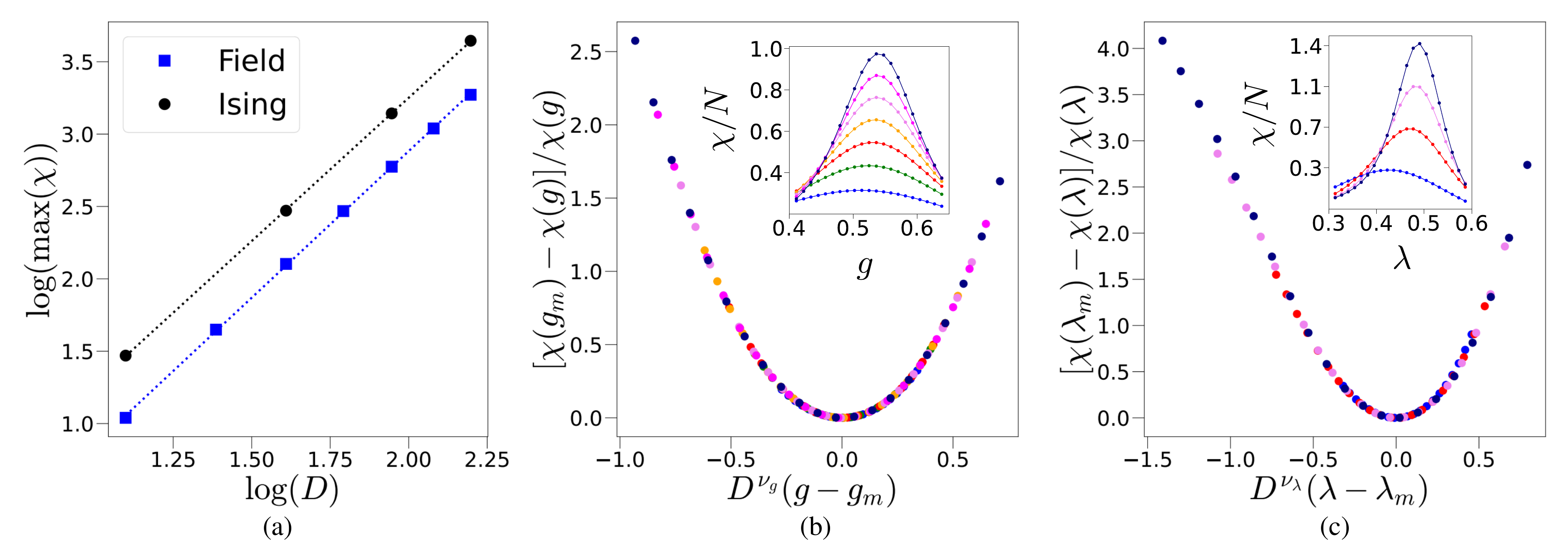}
    \caption{\textbf{Finite-size scaling of fidelity susceptibility for square lattice.} (a) The log - log plot of $\max(\chi)$ with $D$ (Eq.~(\ref{eq:max_chi})) corresponding to the GS of $H$ on a square lattice under the field and the Ising perturbations. The value of slope $k$, same as the exponent appearing in Eq.~(\ref{eq:max_chi}), obtained from the fitting of the data are $k=2.02(1)$ for the field perturbation and $k=1.980(7)$ for the Ising perturbation. Data collapse of $\chi$ obtained by finite size scaling analysis as per Eqs.~(\ref{eq:fs_scaling_field})-(\ref{eq:fs_scaling_ising}) in the case of the field and the Ising perturbations keeping $\lambda=0$ and $g=0$ are shown in (b) and (c) respectively. Insets show the variations of $\chi / N$ as functions of the corresponding perturbation strengths. The values of the exponents (see Eqs.~(\ref{eq:fs_scaling_field})-(\ref{eq:fs_scaling_ising})), obtained from fitting of the data, are $\nu_g=0.900$ and $\nu_\lambda=0.950$, while the other fitting parameters are $\alpha_g=-0.170(3),\delta_g=1.43(2),\alpha_\lambda=-0.601(5)$ and $\delta_\lambda=1.95(1)$. In all the plots, $D\in[3,9]$ and $M=2$, where, only odd values of $D$ are taken for the Ising perturbation (c), and all values of $D\in[3,9]$ are considered in the case of the field perturbation (b). The different lines in the insets of (b) and (c) from bottom to top denote different values of $D$ considered in the increasing order. All the quantities plotted are dimensionless.}
    \label{fig:fidelity}
\end{figure*}

We consider the Kitaev code represented by the Hamiltonian~\cite{kitaev2001,*kitaev2003,*kitaev2006} 
\begin{eqnarray}
    H_K=-\sum_{P}\mathcal{Z}_P-\sum_{V}\mathcal{X}_V,\label{eq:kitaev}
\end{eqnarray}
defined on the square~\cite{kitaev2003,bravyi1998,Trebst2007} and the triangular~\cite{Zarei2015} lattices  with OBC along the vertical ($v$), and PBC along the horizontal ($h$) directions (see Fig.~\ref{fig:kitaev_toric_code})~\cite{bravyi1998,Jamadagni2018}.  
Each of the $z$-type plaquette and $x$-type vertex operators, given by 
\begin{eqnarray}
\mathcal{Z}_P=\bigotimes_{i\in P}\sigma^z_i , \mathcal{X}_V=\bigotimes_{i\in V}\sigma^x_i,
\end{eqnarray} 
follow the commutation relations
\begin{eqnarray}
    [\mathcal{X}_V,\mathcal{Z}_P]=[\mathcal{X}_V,H_K]=[\mathcal{Z}_V,H_K]=0,
\end{eqnarray}
$\forall P,V$ with $P$ and $V$ respectively denoting the plaquette and the vertex indices, and $i$ denoting the indices of the qubits sitting on the edges of the lattice. In the case of the square lattice, both the plaquette and the vertex operators act non-trivially on four qubits in the bulk, while at the boundary, the vertex operators act on three qubits (see Fig.~\ref{fig:kitaev_toric_code}(a)). On the other hand,  in the bulk of the triangular lattice, plaquette and vertex operators act non-trivially on three and six qubits respectively, while at the boundary, the vertex operators have support on $4$ qubits. One can also consider the OBC along $v$ in such a way that the plaquette operators are modified at the open boundary for both the square and the triangular lattices, keeping the vertex operators unaltered~\cite{Jamadagni2018}. However, in this paper, we consider the formalism where plaquette operators remain unchanged. With OBC along $v$, the lattice can be embedded on the surface of a cylinder of height $M\equiv N_{V_v}$ and circumference $D\equiv N_{V_h}$, with $N_{V_h}$ ($N_{V_v}$) being the number of vertices in the horizontal (vertical) directions, and $N_V=D\times M$ being the total number of vertex operators in the system, such that the lattice hosts a total of $N=(2M-1)D$ and $N=(3M-2)D$ qubits in the cases of the square and the triangular lattices respectively. In this paper, we consider a \emph{wide} cylinder in the limit $M\ll D$ (see Fig.~\ref{fig:kitaev_toric_code}(c)), where the open boundary becomes extensive in system size resulting in features that are different from those corresponding to the case of PBC along both directions~\cite{Jamadagni2018}. The constraint 
$\prod_V \mathcal{X}_V = \mathbb{I}$
over the stabilizer operators results in a two-fold degenerate GS manifold $\{ \ket{\Psi_0^\alpha}$, $\alpha=0,1\}$, which are simultaneous eigenstates of all $\mathcal{Z}_P$ and $\mathcal{X}_V$ along with $H_K$. Two non-trivial loops, $L^{x}_{v}$ and $L^z_h$, constitute the loop operators
\begin{eqnarray}
\mathcal{L}^{x}_{v}=\bigotimes_{i\in L^x_v}\sigma^x_i,\; \mathcal{L}^{z}_{h}=\bigotimes_{i\in L^z_h}\sigma^z_i. 
\end{eqnarray}
In terms of the loop and the vertex operators, the GS manifold is given by
\begin{eqnarray}
\ket{\Psi_0^{\alpha}}=\left(\mathcal{L}^x_v\right)^\alpha\left[\prod_{V}\frac{1+\mathcal{X}_V}{2}\right]\ket{0}^{\otimes N},\;\alpha=0,1.
\end{eqnarray}

\begin{figure*}
    \includegraphics[width=0.8\linewidth]{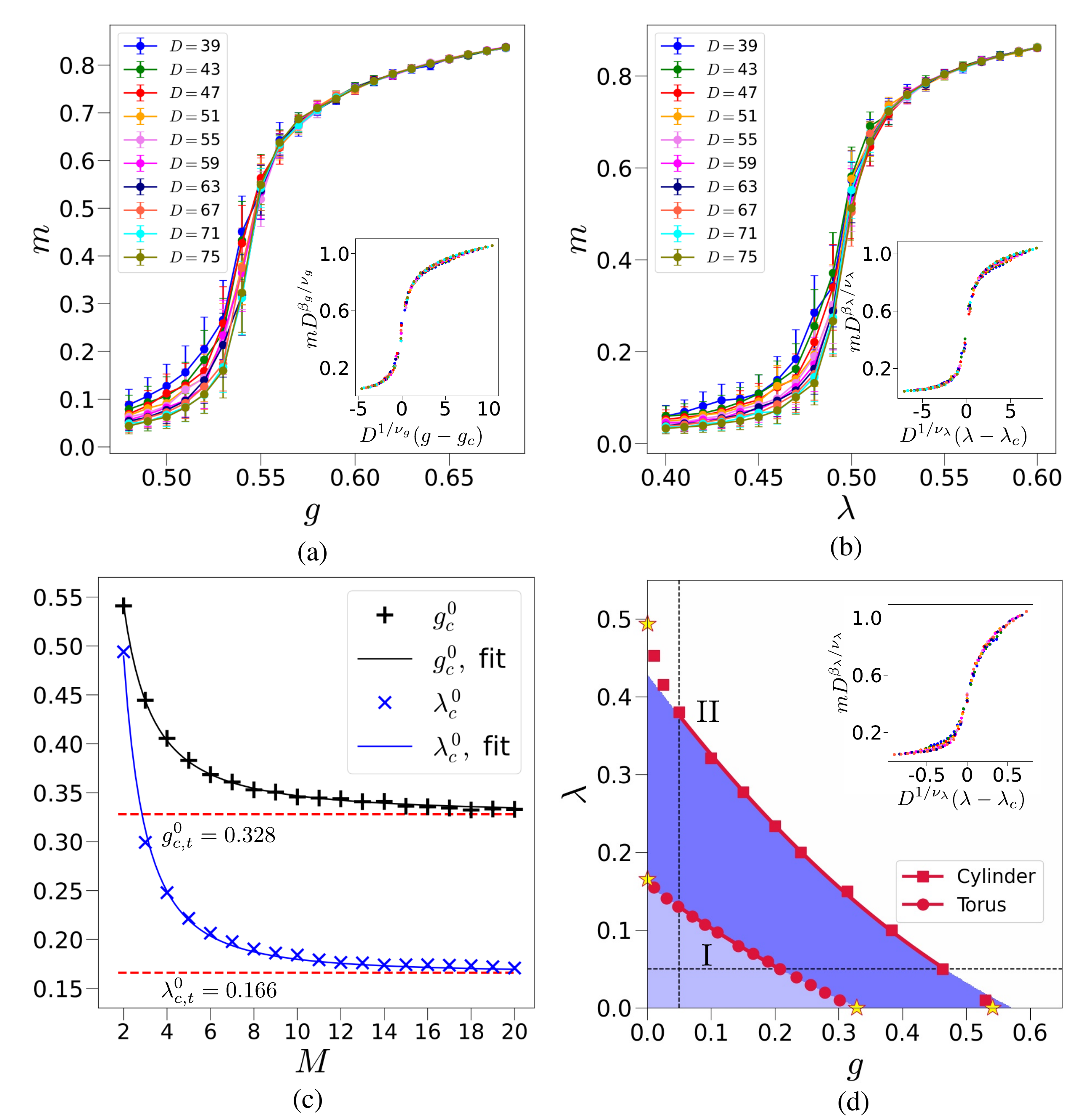}
    \caption{\textbf{Probing QPT with magnetization.} Variation of $m=\langle \tilde{\sigma}_V^z \rangle$ as function of (a) $g$ with $\lambda=0$, and of (b) $\lambda$ with $g=0$ for the square lattice with $M=2$. The insets depict the data collapse of $m$ upon scaling (Eqs.~(\ref{eq:m_scaling_field})-(\ref{eq:m_scaling_ising})), where $\nu_g=\nu_\lambda=1$, $\beta_g=0.163,\beta_\lambda=0.177$, $g_c^0=0.541$ and $\lambda_c^0=0.494$. (c) Variations of $g_c^0$ and $\lambda_c^0$ as functions of $M$, where the QCPs asymptotically approach those corresponding to the 2D square lattice with PBC along both $h$ and $v$ according to Eqs.~(\ref{eq:gc_scaling})-(\ref{eq:lamc_scaling}). The fitting parameters obtained are $\alpha_g=0.608(9), \delta_g=1.50(1), \alpha_\lambda=1.25(4)$ and $\delta_\lambda=1.95(3)$. To support our data, we point out here that the value of $g_c^0$ for $M=3$ obtained in our analysis tends to the previously reported value of $g_c^0=0.453$~\cite{Jamadagni2018} with increasing system size. For $M=2$, we use $D\in[39,75]$ to perform the scaling analysis, while for $M\geq 3$, $g_c^0,\lambda_c^0$ are obtained by the data collapse of $m$ for $11\leq D \leq 25$. (d) The phase boundaries corresponding to the perturbed Kitaev code on the $g-\lambda$ plane, with the inset showing an illustration of the scaling obtained for the case of the phase boundary $\text{I}$ at $g=0.06$. Here, $M=D$ and PBC is assumed along both $h$ and $v$. The phase boundary $\text{II}$ is for the case of $M=2$, PBC along $h$ and OBC along $v$. The boundaries $\text{I}$ and $\text{II}$, given by Eqs.~(\ref{eq:PBC_only})-(\ref{eq:OBC_along_v}) respectively, are obtained by fitting the $(g,\lambda_c^g)$ data acquired from the scaling analysis of $m$ according to Eqs~(\ref{eq:gc_scaling}) and (\ref{eq:lamc_scaling}) with $g,\lambda\geq 5\times 10^{-2}$, and the shaded region exhibiting the enhanced robustness follows the fitted phase boundary. The QCPs corresponding to $(g=0,\lambda>0)$ and $(g>0,\lambda=0)$ are marked by the star attribute. All quantities plotted are dimensionless. 
    }
    \label{fig:magnetization}
\end{figure*}

\subsection{QPT in locally perturbed Kitaev code: Fidelity susceptibility}
\label{subsec:kitaev_qpt_magnetization}

We now consider the locally perturbed Kitaev code embedded on the surface of the wide cylinder ($M\ll D$, see Fig.~\ref{fig:kitaev_toric_code}(c)), and take the square lattice for demonstration. The perturbations are in the form of local fields of strength $g$ on each qubit, and Ising interactions of strength $\lambda$ on each NN qubit-pair, while the full system Hamiltonian reads~\cite{Trebst2007}
\begin{eqnarray}
    H=H_K-g\sum_{i}\sigma^z_i-\lambda\sum_{\langle i,j\rangle}\sigma^z_i\sigma^z_j,
    \label{eq:tqc}
\end{eqnarray}
with $i$, $j$ being qubit-indices, and $\langle i,j\rangle$ representing a NN qubit-pair. In the limit $g=\lambda=0$ ($H=H_K$), the topologically protected two-fold degenerate ground states of the system are genuinely multiparty entangled~\cite{HK2022,HK2023}, while increasing the values $g$ and $\lambda$ drive the state into an unentangled state with no topological order through a QPT from a topological phase to a non-topological phase.

We first probe the QPT using the fidelity 
\begin{eqnarray}
    f(g_1,g_2) &=& \lim_{N\rightarrow\infty} |\left\langle\Psi_0(g_1)|\Psi_0(g_2)\right\rangle|, 
\end{eqnarray}
between the ground states of the model at two different yet close values of a system parameter, say, $g$, given by $g_1$ and $g_2=g_1+\delta$, with $\ket{\Psi_0(g_i)}$ being the GS of the model for $g=g_i,i=1,2$. Fixing $\delta$ and taking different values of $g$, one can compute the second derivative of the fidelity w.r.t. $\delta$, which is referred to as the \emph{fidelity susceptibility} (FS)~\cite{You2007}:
\begin{eqnarray}
    \chi(g) &=& \lim_{\delta\rightarrow 0}\frac{-2\log f(g-\delta/2,g+\delta/2)}{\delta^2}\nonumber\\ &=&-\left.\frac{\partial^2\log f}{\partial\delta^2}\right|_{\delta=0}.
\end{eqnarray}
A peak in the FS for a finite-sized system signals a QPT. A similar definition can also be adopted for FS w.r.t. the Ising interaction strength $\lambda$. 
Our numerical analysis with the GS of $H$ determined via exact diagonalization (ED) indicates that the maximum of FS occurring near the QCP exhibits a power-law divergence as
\begin{equation}\label{eq:max_chi}
    \max(\chi)\propto D^k,
\end{equation} 
which we demonstrate in Fig.~\ref{fig:fidelity}(a). In order to determine the QCPs, we perform the scaling of $\chi$ near the QCPs w.r.t. $g$ and $\lambda$ as (see Fig~\ref{fig:fidelity}(b)-(c))
\begin{eqnarray}
     \lbrack \chi(g_{max})-\chi(g) \rbrack / \chi(g)&=&\chi_g(D^{\nu_g}(g-g_{max})),     \label{eq:fs_scaling_field} \\
    \lbrack \chi(\lambda_{max})-\chi(\lambda) \rbrack / \chi(\lambda) &=& \chi_\lambda (D^{\nu_\lambda}(\lambda-\lambda_{max})),
    \label{eq:fs_scaling_ising}
\end{eqnarray}
respectively, keeping respectively $\lambda$ and $g$ at their zero values. Here,  $\nu_g,\nu_\lambda$ are scaling exponents, $\chi_{g},\chi_{\lambda}$ are chosen to be polynomial functions of $g$ and $\lambda$ respectively~\cite{Gu2008,*WingYu2009,Sandvik_2010}, and $g_{max},\lambda_{max}$ are the points at which FS attains its maximum value for a given system size $D$ and for fixed $M$.
Critical perturbation strengths can be determined by assuming the asymptotic form for $g_{max}$ and $\lambda_{max}$ as,
\begin{eqnarray}
    g_{max} &=& g_c^0+\frac{\alpha_g}{D^{\delta_g}}, \label{eq:fs_gm_finite_size_scaling}\\
    \lambda_{max} &=& \lambda_c^0+\frac{\alpha_\lambda}{D^{\delta_\lambda}}.\label{eq:fs_lam_finite_size_scaling}     
\end{eqnarray}
Here, the superscripts in $g_c^0$ and $\lambda_c^0$ represent the values of $\lambda$ and $g$, respectively, being fixed for determining the QCP. See Appendix~\ref{app:scaling} for details on the procedure for performing the scaling analysis. 

The estimated values of critical perturbation strengths with FS are $g_c^0=0.547,\lambda_c^0=0.496$. Similar scalings are obtained for the case of the triangular lattice as well, where we obtain the QCPs $g_c^0=0.403,\lambda_c^0=0.282$. Note that the propagation of error in the procedure for determining the critical exponents (see Appendix~\ref{app:scaling}) is complex, and we only report the critical exponents up to the third decimal place. On the other hand, the fitting parameters, such as $k$ (Eq.~(\ref{eq:max_chi})), can be estimated via only a least-square fitting of the data, and we include the uncertainty in the estimated value only when the \emph{least} significant digit (which is also the digit with the uncertainty) is within the third decimal place. These critical exponents corresponding to the perturbed Kitaev's toric code model defined on a quasi 1D lattice are not known to the best of our knowledge, and the reported values are purely empirical. We further point out that all values of $D$ used in carrying out the scaling analysis of the FS (see Figure.~\ref{fig:fidelity}) are odd, although in the limit $D\rightarrow\infty$, the QPT occurring  in the system does not depend on whether $D$ is even, or odd. For a discussion on the underlying subtleties behind the choice of odd values of $D$ for carrying out this analysis, see Appendix~\ref{app:odd_even_dichotomy}.

While the FS successfully indicates the QPT, the number of qubits $N=(2M-1)D$ (for square lattice) in the system grows rapidly with $D$ for a fixed $M$, and the QPT quickly becomes intractable using ED. Therefore, to verify the QCPs predicted by the FS with larger system sizes, we consider a mapping of $H$ to  the 2D transverse field Ising model (TFIM) on a zig-zag square lattice with both NN and NNN interactions. Note that $[\mathcal{Z}_P,H]=0$ $\forall P,$ and $H$ is block-diagonal in eigenvalues of $\mathcal{Z}_P$, with a constant energy gap $2$ due to flipping a plaquette operator. For the study of QPT occurring in the GS of $H$, it is sufficient to investigate the low-energy sector of the spectrum where eigenvalues of all $\mathcal{Z}_P$ are $+1$. Note further that $\{\mathcal{X}_V,\sigma^z_i\}=0$ $\forall i\in V$, $[\mathcal{X}_V,\sigma^z_i\sigma^z_j]=0$ if both $i,j\in V$, and $\{\mathcal{X}_V,\sigma^z_{i}\sigma^z_j\}=0$ if either $i$ or $j$ $\in V$, allowing one to switch from the vertex operators to an effective spin language where $\pm 1$ eigenvalues of $\mathcal{X}_V$ serve as the basic degrees of freedom, and introduce a pseudo-qubit operator $\tilde{\sigma}^x_V$ for $\mathcal{X}_V$. This maps $H$ to the effective 2D TFIM, represented by~\cite{Trebst2007,Hamma2008,Wu2012,Zarei2015,Kogut1979,*Wegner2014}
\begin{eqnarray}
    \tilde{H}=-g\sum_{\langle V,V^\prime\rangle}\tilde{\sigma}^z_V\tilde{\sigma}^z_{V^\prime}-2\lambda\sum_{\langle V,V^{\prime\prime}\rangle}\tilde{\sigma}^z_V\tilde{\sigma}^z_{V^{\prime\prime}} -\sum_{V}\tilde{\sigma}^x_V,
    \label{eq:effective_ising_square}
\end{eqnarray}
on a zig-zag square lattice (see Fig.~\ref{fig:kitaev_toric_code}(a)) of size $N_V$ with PBC (OBC) along $h$ ($v$),  where the pseudo-qubits sit on the vertices $V$ of the original lattice, and $\langle V,V^\prime\rangle$ and $\langle V,V^{\prime\prime}\rangle$ respectively represents the NN and NNN effective qubit-pairs with interaction strengths $g$ and $2\lambda$ respectively.

\begin{figure*}
    \includegraphics[width=0.9\linewidth]{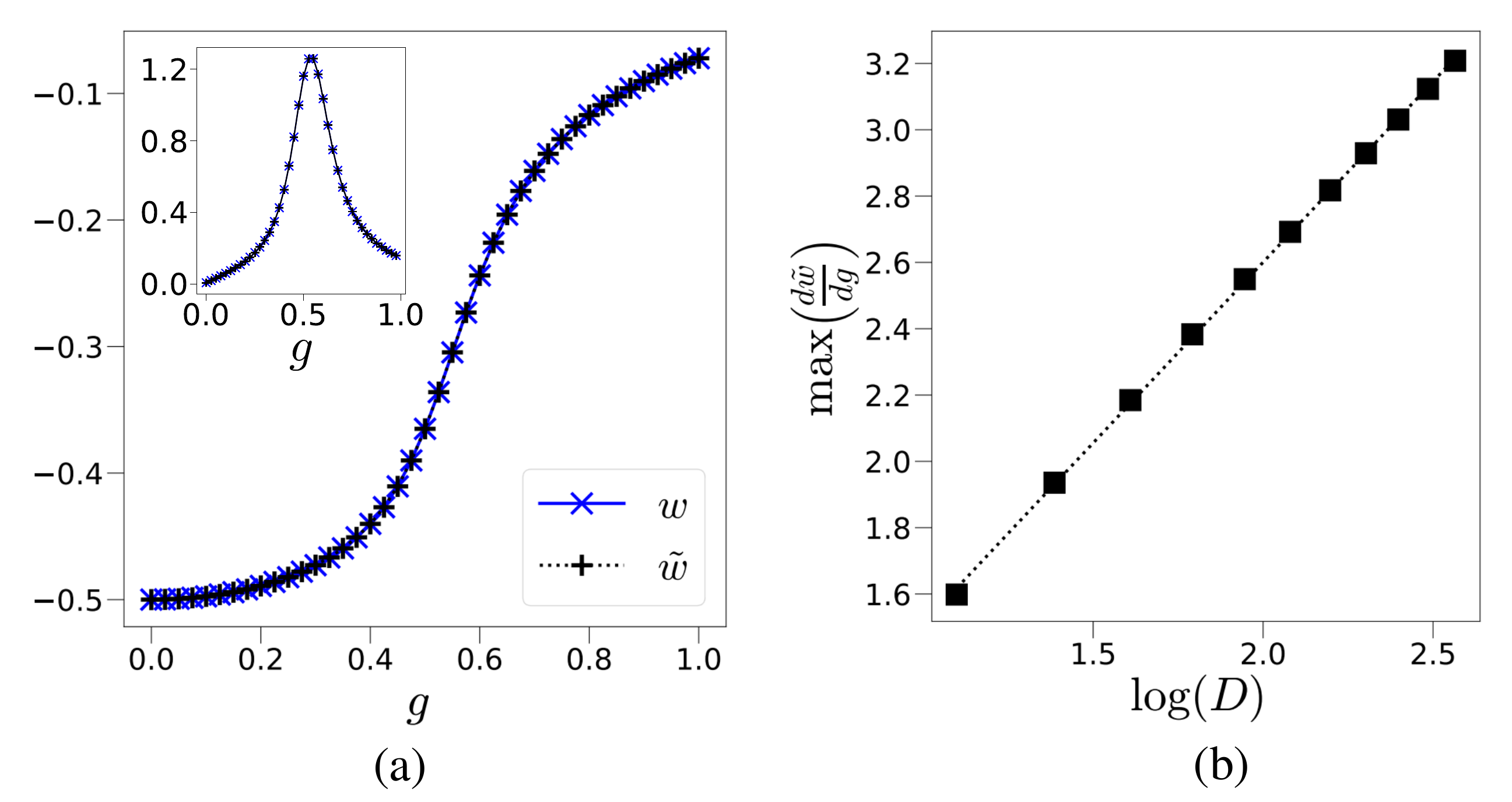}
    \caption{\textbf{Expectation value of local witness operator.} (a) Variation of $\tilde{w}$ and (inset) $d\tilde{w}/dg$ as a function of $g$ for the lattice size $D=7,M=2$ $(N=21)$.  (b) Variation of $\max({d\tilde{w}/ dg}$) against system size $\log(D)$ according to Eq.~(\ref{eq:scaling_maxwit_logD}) with $M=2,$ and $D\in [3,13]$. The value of $k$ (see Eq.~(\ref{eq:scaling_maxwit_logD})) is found to be $1.090(6)$. All quantities plotted are dimensionless.
    }
    \label{fig:comparison}
\end{figure*}

For demonstration, we choose the \textit{Kitaev ladder} with $M=2$, and fix $D$ to be odd. For the corresponding $\tilde{H}$, we compute $m=\langle\tilde{\sigma}^z_V\rangle$ using continuous time quantum Monte Carlo (QMC) simulation with cluster updates~\cite{Rieger1999}. The QPT is captured by a  scaling of $m$ as
\begin{eqnarray}
    m&=&D^{-\beta_g / \nu_g}m_g\left[D^{1/ \nu_g}\left(g-g_c^0\right)\right],    \label{eq:m_scaling_field} \\
    m&=&D^{-\beta_\lambda / \nu_\lambda}m_\lambda\left[D^{1 / \nu_\lambda}\left(\lambda-\lambda_c^0\right)\right]        
    \label{eq:m_scaling_ising}
\end{eqnarray}
with $\beta_g (\beta_\lambda)$ and $\nu_g (\nu_\lambda)$ being the critical exponents corresponding to the field (Ising) perturbation, and $m_{g} (m_\lambda)$ is a polynomial in $g$ ($\lambda$)~\cite{Sandvik_2010}. 
In Fig.~\ref{fig:magnetization}(a)-(b) we demonstrate the scaling in the case of square lattice under field and Ising perturbations. Similar to the FS, we determine the critical exponents to be $\nu_g=\nu_\lambda=1$, and obtain $\beta_g=0.163$ $\beta_\lambda=0.177$. Further the critical values of the field and the Ising interaction strengths are found to be   $g_c^0=0.541$ and $\lambda_c^0=0.494$, which are in agreement with the ones obtained from the FS analysis (see also Appendix~\ref{app:odd_even_dichotomy}).

To explore how the QCPs shift with an increase in the system size by increasing $M$ in Fig.~\ref{fig:magnetization}(c), we plot the variations of $g_c^0$ and $\lambda_c^0$ with $M$, where for each value of $M\geq 3$, the QCPs are obtained by the data collapse of $m$ for $11\leq D \leq 25$, and for $M=2$, by that of system sizes $39\leq D \leq 75$ (see Table~\ref{tab:qpt} and Fig.~\ref{fig:magnetization}). The QCPs asymptotically approach their corresponding values for the Kitaev code on a square lattice with PBC along both directions as,
\begin{eqnarray}
    g_c^0=g_{c,t}^0+\frac{\alpha_g}{M^{\delta_g}}\label{eq:gc_scaling},\\
    \lambda_c^0=\lambda_{c,t}^0+\frac{\alpha_\lambda}{M^{\delta_\lambda}}\label{eq:lamc_scaling}.
\end{eqnarray}

Determination of the critical points $\lambda_c^g$ for non zero values of $g$ via scaling analysis of $m$, and subsequent fitting of the data $(g,\lambda_c^g)$ lead to the phase boundaries 
\begin{eqnarray}
    \text{I: }\lambda &=&0.158-0.59(1) g +0.35(4) g^2,    \label{eq:PBC_only}\\
    \text{II: } \lambda &=& 0.429(4)-1.09(4) g +0.59(7) g^2,\label{eq:OBC_along_v}
\end{eqnarray}
with Eq.~(\ref{eq:PBC_only}) (Eq.~(\ref{eq:OBC_along_v})) corresponding to PBC along both $h$ and $v$ (PBC along $h$, OBC along $v$) (see Fig.~\ref{fig:magnetization}(d)), where data corresponding to $g,\lambda \geq 5\times 10^{-2}$ is used for the fitting.
This is motivated from the fact that the mapping of the Kitaev model on a cylinder to a 2D Ising model leads to different coordination numbers of the mapped lattice for the cases of $g,\lambda>0$, $(g=0,\lambda>0)$, and $(g>0,\lambda=0)$. More specifically, for $M=2$, the coordination numbers are (a) $5$ for $(g,\lambda>0)$, (b) 2 for  $(g=0,\lambda>0)$, and (c) 3 for $(g>0,\lambda=0)$, while for $M>2$,  the effective model has additional qubits in the bulk, with their coordination numbers (a) 4 in the limits $(g>0,\lambda=0)$ and $(g=0,\lambda>0)$, and (b) 8 when $\lambda,g>0$. As the critical points are strongly dependent on the coordination number of the lattice, in the limit $\lambda,g\rightarrow 0$ for the case of square lattice, the critical point shifts relatively differently as compared to the case when both $\lambda,g>0$. This is clearly visible from the data in the range $g<5\times 10^{-2}$ shown in Fig.~\ref{fig:magnetization}(d) (the QCP corresponding to $g=0$ is denoted by the star attribute), which clearly demonstrates a different trend compared to the fitted quadratic phase boundary for $g,\lambda>0.05$. It is clear from the figure that the topological phase of the Kitaev code with OBC along $v$, for $g,\lambda>0$, survives considerably larger perturbations in terms of $g$ and $\lambda$, and therefore exhibits a higher robustness as compared to the Kitaev code with similar perturbations defined on a square lattice with PBC along both directions. Further, the robustness increases as one reduces the value of $M$, i.e, as one approaches the $M\ll D$ limit, as clearly seen from Fig.~\ref{fig:magnetization}(c). 

Similar treatment of the Kitaev code on the triangular lattice (see Fig.~\ref{fig:kitaev_toric_code}(b)) with OBC along the vertical direction maps the model to the TFIM on the triangular lattice having only NN interactions, but different NN interaction strengths in the boundary and the bulk (see Fig.~\ref{fig:kitaev_toric_code}(b)):
\begin{eqnarray}
    \tilde{H}&=&-\sum_{V}\tilde{\sigma}^x_V-(g+2\lambda)\sum_{\langle V,V^\prime\rangle\in\text{bulk}}\tilde{\sigma}^z_V\tilde{\sigma}^z_{V^\prime} \nonumber\\
    &&-(g+\lambda)\sum_{\langle V,V^\prime\rangle\in\text{boundary}}\tilde{\sigma}^z_V\tilde{\sigma}^z_{V^\prime}.
    \label{eq:effective_ising_triangular}
\end{eqnarray}
The QPT in the model can be probed in a similar fashion as in the case of the square lattice, while all features of the FS and magnetization scaling in the vicinity of the QPT, and the behaviour of critical strengths $g_c^0,\lambda_c^0$ against $M$ remain qualitatively similar. Note that in contrast with the parabolic phase boundary obtained in the case of the square lattice, the phase boundary in the present case is a straight line, given by
\begin{eqnarray}
    \text{I: }\lambda &=&0.103-0.498(4)g,\\
    \text{II: }\lambda &=&0.279(1)-0.693(4)g,
\end{eqnarray}
with $g,\lambda \geq 10^{-2}$. Further, the effective Ising model obtained from the Kitaev code on the triangular lattice always contains boundary qubits (bulk qubits) with a coordination number four (six) irrespective of value of $\lambda$ and $g$, and therefore, the boundary effect close to $(g=0,\lambda>0)$ and $(g>0,\lambda=0)$ is absent in the case of the Kitaev code on the triangular lattice. 

In Table~\ref{tab:qpt}, we list the known critical perturbation strengths for the topological-to-non-topological QPTs in the case of Kitaev code on the square and the triangular lattices with different boundary conditions, along with the QCPs determined in this paper. For determination of these QCPs, we use FS as well as magnetization of the mapped 2D Ising model calculated using  the QMC technique as the order parameter. From our investigation, it is clear that a system size extensive open boundary results in a significant enhancement in the robustness of the Kitaev code under local perturbations for both square and triangular lattices. We also point out that in the limit $M\gg D$ of the \emph{narrow} cylinder, the bulk of the system for both square and triangular lattices behaves similar to the case of PBC along both directions, with small increase in $g_c^0$ and $\lambda_c^0$ from their respective values corresponding to PBC along both $h$ and $v$ (see Table \ref{tab:qpt}).

\subsection{Probing the QPT with entanglement}
\label{subsec:kitaev_entanglement}

\begin{figure*}
    \centering
    \includegraphics[width=0.9\linewidth]{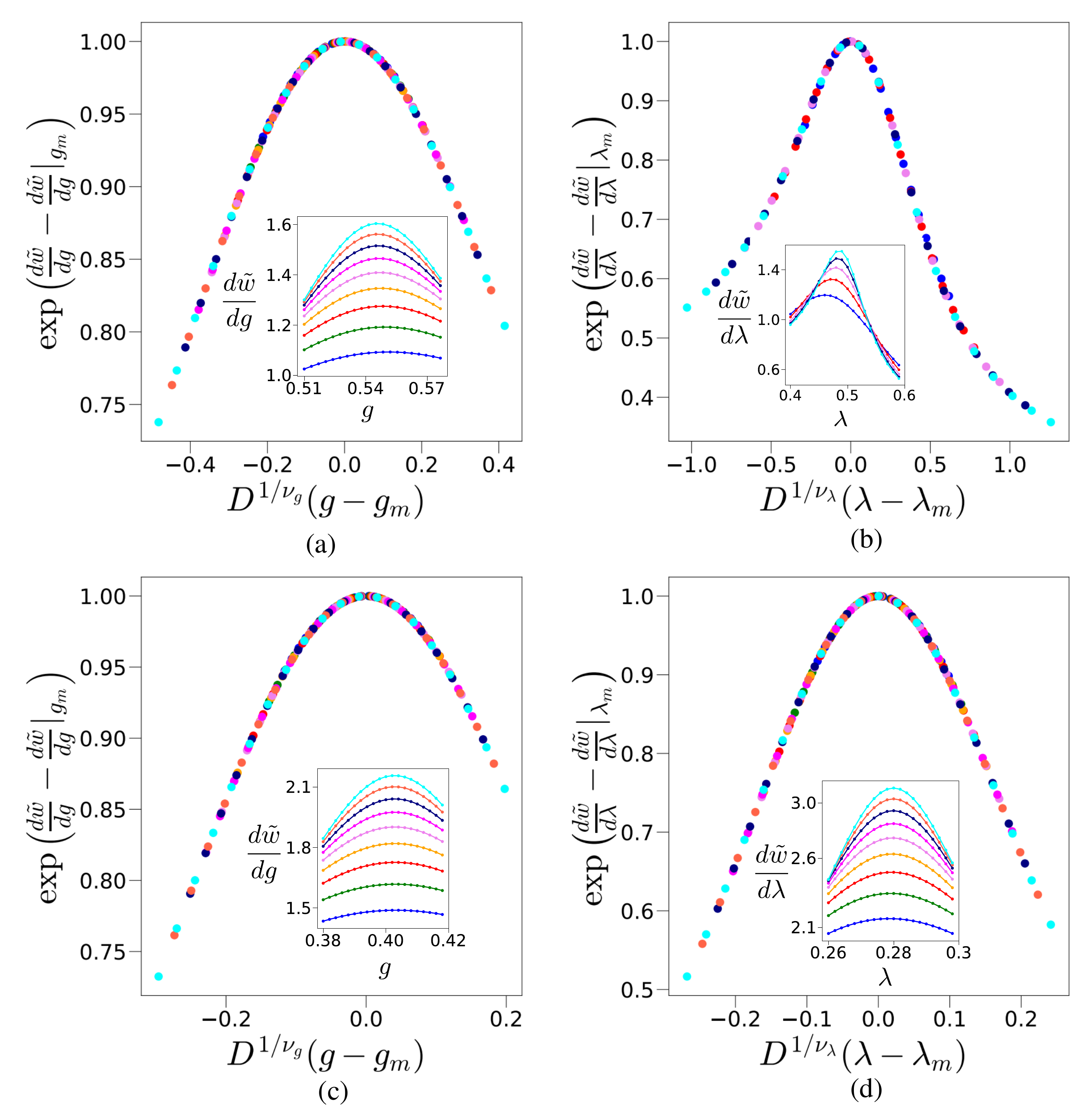}
    \caption{\textbf{Finite-size scaling of $d\tilde{w}/ dg$ for square and triangular lattices.} Data collapse obtained by the finite size scaling as per Eq.~(\ref{eq:witness_g_finite_size_scaling}) for the case of (a)-(b) square lattice and (c)-(d) triangular lattice. Insets show the variations of the first derivative of $\tilde{w}$ w.r.t. the system parameter for various system sizes near the QCP. All plots correspond to system sizes $M=2$ and $D\in [5,13]$, where only odd values of $D$ for the square lattice under Ising perturbation (b) are considered (similar to Fig.~\ref{fig:fidelity}). The critical exponents are $\nu_g=0.985$ $(\nu_g=1)$ and $\nu_\lambda=1$ $(\nu_\lambda=0.988)$ for the square (triangular) lattice. Other fitting parameters, obtained by fitting the data to Eq.~(\ref{eq:witness_gm_finite_size_scaling}) are $\alpha_g=0.077$ $(\alpha_g=0.009),$ $\delta_g=1.457(6)$ $(\delta_g=1.77(1)),$ $\alpha_\lambda= -0.244(4)$ $(\alpha_\lambda=-0.018)$ and $\delta_\lambda=1.16(1)$ $(\delta_\lambda=1.47(1))$ in the case of the square (triangular) lattice. The different lines in the insets of all figures from bottom to top denote different values of $D$ considered in the increasing order. All quantities plotted are dimensionless.
    }
    \label{fig:witness_scaling}
\end{figure*}

Computing entanglement in the GS of $H$ is hindered by the requirement of full access to the GS of the model, which, in turn, is restricted by the exponential growth of the Hilbert space of the system with the system-size. The mapping $H\rightarrow\tilde{H}$ results in the loss of information regarding the degeneracy and the entanglement-content of the GS of the perturbed model. Nevertheless, we show that the mapping can be exploited to probe entanglement of the locally perturbed Kitaev code with OBC along $v$, which can subsequently be used to study the topological-to-non-topological QPT occurring in $H$. We choose the square lattice for demonstration, and use information available on the two-fold degenerate TPGSs of $H_K$ ($H$ in the limit $g=\lambda=0$~\cite{kitaev2003,bravyi1998,Jamadagni2018}), which is known to facilitates non-zero LE~\cite{popp2005,verstraete2004,HK2022,HK2023} over the non-trivial loop $L_{v}^{x}$ via  $\sigma^z$ measurements on all qubits belonging not to $L_{v}^{x}$ but to the plaquettes through which $L_{v}^{x}$ passes, and $\sigma^x$ measurements on all other qubits~\cite{HK2022,HK2023}\footnote{Note that one can also choose the horizontal non-trivial loop as the subsystem. However, we choose the vertical loop for this paper to keep the subsystem-size constant with increasing system size governed by $D$.} (see Appendix~\ref{app:localizable_entanglement} for details). A witness-based lower bound to the LE over $L_v^x$ can be obtained as a function of the expectation value
\begin{equation}
    w=\text{Tr}(W\rho)
\end{equation}
of the local witness operator $W$, constructed as~\cite{Alba2010,Amaro2020}   
\begin{eqnarray}
    W=\frac{1}{2}-\prod_{\mathcal{S}_{\alpha^\prime}\in\left\{\mathcal{S}_{\alpha}\right\}}\frac{I+\mathcal{S}_{\alpha^\prime}}{2}
    \label{eq:witness}
\end{eqnarray}
in a TPGS $\rho$ of (\ref{eq:tqc}). Here, $\{\mathcal{S}_\alpha\}$ is a subset of size $n$ of all possible stabilizers that can be constructed using the plaquette and the vertex stabilizers. Note that for the square (triangular) lattice, $n=M$ $(n=2M-1)$ for $L_v^x$, as shown in Fig.~\ref{fig:kitaev_toric_code}. The set $\{\mathcal{S}_\alpha\}$ is chosen such that the Pauli matrices contributing to $\mathcal{S}_\alpha$ must commute outside the chosen subset $L_{v}^x$, and the components of $\mathcal{S}_\alpha$ on the chosen subset must themselves form a complete set of stabilizer generators corresponding to a stabilizer state on the chosen qubits (see Appendix~\ref{app:witness_construction} for a detailed discussion on the construction of the witness operator $W$)~\cite{Amaro2020}. For example, in the case of $L^{x}_v$ in the Kitaev code on square lattice with OBC along $v$, $\{\mathcal{S}_\alpha\}$ is constituted of all the plaquette stabilizers through which $L_v^x$ passes, and the subset of the vertex stabilizers that have a common support with $L^x_v$ in terms of a single qubit (see Appendix~\ref{app:witness_construction} and Fig.~(\ref{fig:witness_composition}))~\cite{HK2022}. Using genuine multiparty entanglement~\cite{horodecki2009} quantified by geometric measures~\cite{Shimony1995,*Barnum2001,*Wei2003}, one can determine a lower bound for the localizable genuine multiparty entanglement~\cite{HK2023} on $L_v^x$ in terms of $w$ as~\cite{guhne2007} (also see Appendix~\ref{app:GM_witness_lb} for details)
\begin{equation}
    E^w=\frac{1-\sqrt{1-4w^2}}{2},
    \label{eq:lb_gme}
\end{equation}
while choosing negativity~\cite{peres1996,*horodecki1996,*zyczkowski1998,*vidal2002} as a bipartite entanglement measure, the lower bound for localizable bipartite entanglement between a boundary qubit and the rest of the qubits in $L^{x}_{v}$ is given by~\cite{HK2022},
\begin{equation}
    E^w=-2w \label{eq:witness_lowerbound}.
\end{equation}
We point out here that the witness-based lower bound corresponding to the LE computed in terms of negativity provides the actual value of LE for the TPGS of $H_K (g=0,\lambda=0)$~\cite{HK2022}. Using this lower bound, in~\cite{HK2022}, it has been shown for finite-sized systems that the first derivative of the lower bound potentially capture a signature of the QPT for $g>0$, with $\lambda=0$ and PBC along both $h$ and $v$. However, a complete scaling analysis for entanglement in the case of non-vanishing perturbation strengths in the limit $D\rightarrow\infty$ is a non trivial task. 

Noting that one can access the lower bound of LE by computing $w$, in this paper, we probe the QPT in terms of $w$. Without any loss in generality, the construction of $\{\mathcal{S}_\alpha\}$ for $L^x_{v}$ is carried out ensuring $\mathcal{S}_{\alpha=1}$ having the form $\mathcal{S}_1=\prod_{V\in L_v^x}\mathcal{X}_V$, and the rest $n-1$ stabilizers being made of plaquette stabilizers only as $\mathcal{S}_\alpha=\bigotimes_{P\in L_v^x}\mathcal{Z}_{P}$, $\alpha=2,\cdots,n$ (see Appendix~\ref{app:witness_construction} for the details of construction of $W$ along $L_v^x$ considered in this paper). This structure of $\{\mathcal{S}_\alpha\}$ along with the fact that $\mathcal{Z}_P=+1\forall P$ in the GS of $H$ (Eq.~(\ref{eq:tqc})) simplifies $w$ as $w=-\langle \mathcal{S}_1 \rangle/2$, which is equivalent to computing
\begin{equation}
    \tilde{w}=-\frac{\langle\tilde{\mathcal{S}}_1\rangle}{2}
\end{equation}
in the GS of TFIM, where we have used the mapping $\mathcal{X}_V\rightarrow\tilde{\sigma}^x_V$ to obtain, 
\begin{eqnarray}    \tilde{\mathcal{S}}_1=\prod_{i\forall\mathcal{X}_{V_i}\in \mathcal{S}_1}\tilde{\sigma}^x_i.
\end{eqnarray}  
Note that the ``tilde" symbol distinguishes the expectation value of the witness operator in the GS of the mapped TFIM from the same computed from the GS of the perturbed Kitaev model. This provides an advantage in computing $w$ in terms of $\tilde{w}$ using ED, since the effective lattice of size $M\times D$ corresponding to $\tilde{H}$ stands for the locally perturbed Kitaev code on a square lattice hosting $N=(2M-1)D > MD$ qubits.

In Fig.~\ref{fig:comparison}(a), we plot $w$ and  $\tilde{w}$ as functions of $g$ for the Kitaev code on a square lattice ($D=7$, $M=2$), and show them to be identical. The absolute value of the first derivative of $\tilde{w}$ w.r.t. $g$ is plotted in the inset as a function of $g$, exhibiting a sharp peak near the QCP $g_c^0$.  Fig.~\ref{fig:comparison}(b) demonstrates how the maximum of $d\tilde{w}/dg$ approaches infinity with increasing system-size via increasing $D$:
\begin{eqnarray}
    \max{\left( \frac{d \tilde{w}}{dg}  \right)} = k\log{(D)}+\text{constant}\label{eq:scaling_maxwit_logD}.
\end{eqnarray}
We point out the similarity with the logarithmic divergence observed in the NN concurrence in the case of the 1D TFIM~\cite{osterloh2002}. Fitting the data with Eq.~(\ref{eq:scaling_maxwit_logD}) leads to $k=1.09(0)$, for the case reported in Fig.~\ref{fig:comparison}(b) while similar scaling (Eq.~(\ref{eq:scaling_maxwit_logD})) is observed in other cases (eg. Ising perturbation in the square lattice, or field and Ising perturbation in the triangular lattice) as well except different scaling exponent $k$. Since the logarithmic divergence occurs in the limit $D\rightarrow\infty$, in order to obtain the QCP $g_c^0$, we perform the scaling~\cite{osterloh2002} 
\begin{equation}
    \exp \left( \frac{d \tilde{w}}{dg}-\frac{d \tilde{w}}{dg}\bigg|_{g=g_{max}} \right)=F_g(D^{1/ \nu_g}(g-g_{max}))\label{eq:witness_g_finite_size_scaling},
\end{equation}
where $\nu_g$ is the critical exponent, $g_{max}$ is the value of $g$ at which the maximum in $d\tilde{w}/dg$ takes place for a given system size $D$, and $F_g$ is a polynomial scaling function. The QCP, $g_c^0$ is obtained by assuming the asymptotic form for $g_{max}$ (similar to Eqs.~(\ref{eq:fs_gm_finite_size_scaling})-(\ref{eq:fs_lam_finite_size_scaling})):
\begin{equation}
    g_{max}=g_c^0+\frac{\alpha_g}{D^{\delta_g}}\label{eq:witness_gm_finite_size_scaling}
\end{equation}
where $\alpha_g,\delta_g$ are two fitting parameters specific for each type of lattice. Similar scaling forms are chosen for the Ising perturbation as well. Figs.~\ref{fig:witness_scaling}(a)-(b) show the data collapse in the case of square lattice
with the QCPs $g_c^0=0.543$ and $\lambda_c^0=0.497$ and Qualitatively similar features are observed on the triangular lattice also (see Figs.~\ref{fig:witness_scaling}(c)-(d)), with the QCPs $g_c^0=0.402$ and $\lambda_c^0=0.280$. The results are in agreement with the critical strengths obtained earlier by the analysis of FS and magnetization.

\begin{figure*}
\includegraphics[width=0.9\linewidth]{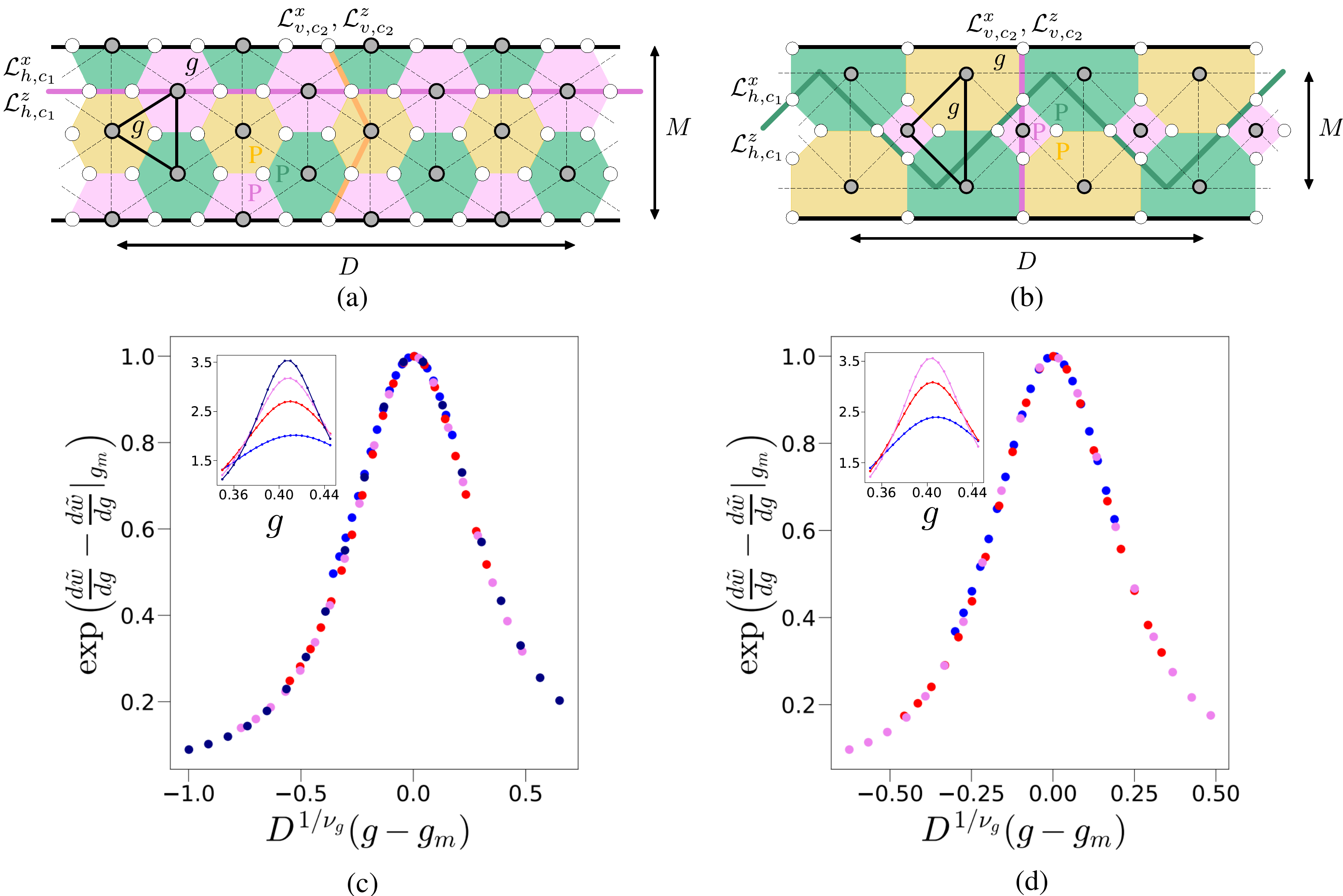}
\caption{Color code defined on the (a) honeycomb and the (b) square-octagonal lattices with open (periodic) boundary condition along the vertical (horizontal) direction, where qubits (denoted by white circles) sit on the vertices of the lattice. The honeycomb lattice involves $6~(4)$ qubits in the bulk (boundary), while the square-octagonal lattice involves $4~(6)$ qubits in the bulk (boundary). The horizontal and the vertical non-trivial loops corresponding to logical operators $\mathcal{L}^{x(z)}_{h,c_1}$ and $\mathcal{L}^{x(z)}_{v,c_2}$, are shown with thick colored lines, where $h$ and $v$ denote the horizontal and vertical directions, and $c_1,c_2$ denote two different colors. The horizontal length of the lattice is assumed to be $D$ and vertical length $M$, with $D\gg M$. The dashed thin black lines represent the lattice on which effective qubits (denoted with gray circles) are located. The data collapse obtained for the derivative of $\tilde{w}$ in the case of the (c) honeycomb lattice with $D\in[4,10]$, and in the case of the (d) square-octagonal lattice with $D\in[4,8]$. Insets in (c) and (d) show the variations of $d \tilde{w}/dg$ w.r.t. $g$ in the respective cases. The data collapse occurs at the critical perturbation strength $g_c^0=0.406~(g_c^0=0.401) $ and the exponent $\nu_g=0.806~(\nu_g=0.846)$ for the honeycomb (square-octagonal) lattice. 
Other fitting parameters (Eq.~(\ref{eq:witness_gm_finite_size_scaling})) are $\alpha_g=0.136(9)~(\alpha_g=-1.627)$ and $\delta_g=2.09(5)~(\delta_g=0.067)$ for the honeycomb (square-octagonal) lattice. The different lines in the insets of (c) and (d) from bottom to top denote different values of $D$ considered in the increasing order. All quantities plotted are dimensionless.
}
\label{fig:color_code}
\end{figure*}

\section{Locally perturbed color code}
\label{sec:color}

We now focus on the color code defined on tricolorable lattices~\cite{bombin2006,bombin2007}. The Hamiltonian of the color code is given by
\begin{eqnarray}
    H_C=-\sum_{P}\mathcal{Z}_P-\sum_{P}\mathcal{X}_P,\label{eq:color},
\end{eqnarray}
where every plaquette $P$ is associated with both $x$ and $z$ type plaquette operators,
\begin{eqnarray}
\mathcal{Z}_P=\bigotimes_{i\in P}\sigma^z_i;\;\;\mathcal{X}_P=\bigotimes_{i\in P}\sigma^x_i,
\end{eqnarray}
with $i\in P$ denoting all qubits belonging to the plaquette $P$. Note that unlike the Kitaev code, qubits in the color code are placed on the vertices of the lattice. The plaquette operators form a mutually commuting stabilizer set, i.e,
\begin{eqnarray}
    [\mathcal{X}_P,\mathcal{Z}_P]=[\mathcal{X}_P,H_C]=[\mathcal{Z}_P,H_C]=0.
\end{eqnarray}
While the QPT occuring in the color code with PBC along both horizontal and vertical directions under parallel magnetic field and Ising interactions have been studied~\cite{Jahromi2013,Zarei2015,Jahromi2013a}, the same with open boundaries are yet to be explored. We consider the color code defined on the honeycomb and the square-octagonal lattices~\cite{Landahl2011} with OBC along the vertical, and PBC along horizontal direction (see Figs.~\ref{fig:color_code}(a)-(b) respectively). The honeycomb lattice contains plaquette operators acting non-trivially on six qubits in the bulk and four qubits in the boundary, while the square-octagonal lattice contains plaquette operators acting non-trivially on four and eight qubits in the bulk, and six qubits in the boundary. We denote the maximum number of plaquettes appearing in the vertical direction to be $M$ and the same along the horizontal direction by $D$. The lattice hosts a total of $N=D(2M-1)$ qubits for the honeycomb lattice, and $N=2D(2M-1)$ qubits for the square-octagonal lattice respectively, with only even values of $D$ allowed in order to simultaneously maintain tricolorability of the lattice and PBC along $h$. We choose the smallest possible instance of $M$—$M=3$ for the honeycomb lattice, and $M=2$ for the square-octagonal lattice—required for degenerate TPGS for investigating the QPT in the extensive open boundary limit by increasing $D$. The TPGS manifold, $\{\ket{\Psi^{\alpha,\beta}_0},\alpha,\beta=0,1 \}$, is fourfold degenerate, and four independent non-trivial loop operators, $\mathcal{L}_{h(v),c_1(c2)}^{x(z)}$—two of each of the $x$ and the $z$ type—can be identified along vertical and horizontal directions acting non-trivially on the qubits lying along the path $L_{h}^{x(z)}$ and $L_{v}^{x(z)}$ respectively. Here, $c_1,c_2$ denotes two different colors as shown in Fig.~\ref{fig:color_code}(a)-(b). The $\mathcal{L}^x_h$ and $\mathcal{L}^x_v$ operators can be used to span the GS manifold of $H_C$ as follows:
\begin{eqnarray}
\ket{\Psi_0^{\alpha,\beta}}=\left( \mathcal{L}^x_{h,c1}\right)^\alpha \left( \mathcal{L}^x_{v,c2}\right)^\beta \left[\prod_{P}\frac{1+\mathcal{X}_P}{2}\right]\ket{0}^{\otimes N}.
\end{eqnarray}

We now consider the color code with OBC along $v$, perturbed by a parallel magnetic field:
\begin{eqnarray}
    H=H_C-g\sum_{i}\sigma^z_i.
    \label{eq:tqc_color}
\end{eqnarray}
The actual model becomes very quickly intractable with ED even for the smallest possible values of $M$ and the analysis with FS becomes inconclusive due to the inaccessibility of the full GS of the perturbed model. However, the perturbed color code with OBC along $v$, given by Eq.~(\ref{eq:tqc_color}), can be mapped to the Baxter-Wu model in a transverse field with additional Ising interactions along the boundary (see~\cite{Jahromi2013,Jahromi2013a,Capponi2014} for a similar mapping of the perturbed color code with PBC along both $h$ and $v$) as follows. Similar to the Kitaev code, $[H,\mathcal{Z}_P]=0\forall P$, and for the QPT, the GS of $H$ is guaranteed to have an eigenvalue of $+1\forall \mathcal{Z}_P$. Further, $\{ \sigma_i^z,\mathcal{X}_P \} \forall i\in P$ allows one to treat the $\mathcal{X}_P$ operators as effective qubits, and introduce the pseudo spin operator $\tilde{\sigma}^x_P$ residing on every plaquette with two eigenvalues $\pm 1$ that are flipped by the field term. Since the lattice is tricolorable and qubits reside on the vertices of the lattice, the field term on every qubit in the bulk results in a three-body Baxter-Wu interaction $\tilde{\sigma}_P^z\tilde{\sigma}_{P^\prime}^z\tilde{\sigma}_{P^{\prime\prime}}^z$ between the neighbouring effective qubits. On the other hand, qubits in the boundary belong to only two plaquettes, resulting in an Ising term $\tilde{\sigma}_P^z\tilde{\sigma}_{P^\prime}^z$ between the neighbouring effective qubits lying along the boundary. Thus, the effective Hamiltonian reads
\begin{eqnarray}
    \tilde{H}&=&-\sum_{P}\tilde{\sigma}^x_P-g\sum_{\langle P,P^\prime,P^{\prime\prime}\rangle\in\text{bulk}}\tilde{\sigma}^z_P\tilde{\sigma}^z_{P^\prime}\tilde{\sigma}^z_{P^{\prime\prime}} \nonumber\\
    &&-g\sum_{\langle P,P^\prime\rangle\in\text{boundary}}\tilde{\sigma}^z_P\tilde{\sigma}^z_{P^\prime}.
    \label{eq:effective_color}
\end{eqnarray}
The lattices in which the effective qubits reside are shown using dashed lines in Figs~\ref{fig:color_code}(a)-(b). 

Similar to the Kitaev code, we focus on the expectation value of $w$ of the local witness operator constructed for obtaining a lower bound of LE on $L_v^x$, and compute $\tilde{w}$ in the GS of the effective model $\tilde{H}$. The scaling of $\max(d\tilde{w}/ dg)$ with $\log(D)$ is found to be the same as Eq.~(\ref{eq:scaling_maxwit_logD}), with $k=3.30(3)$ ($k=3.34(2)$) for the honeycomb (square-octagonal) lattice, which are different from that of Kitaev code. Further, the data collapse obtained for $d\tilde{w}/ dg$ as per Eq.~(\ref{eq:witness_g_finite_size_scaling}) for the honeycomb and the square-octagonal lattices are demonstrated in Figs.~\ref{fig:color_code}(c) and \ref{fig:color_code}(d) respectively. Our analysis provides QCPs as $g_c^0=0.406$ for the honeycomb lattice, and $g_c^0=0.401$ for the square-octagonal lattice. We point out here that $g_c^0$ for the honeycomb lattice with OBC along $v$ is close to the known QCP $g_c^0=0.383$ corresponding to the perturbed color code on the honeycomb lattice with PBC along both $h$ and $v$~\cite{Jahromi2013,Jahromi2013a}. This indicates a lower enhancement in the topological phase in the case of the color code against perturbation in the form of a parallel field when the boundary is made open along $v$. 

\section{Conclusion and outlook}
\label{sec:conclusion}

In this paper, we study the Kitaev code in the presence of local magnetic field and spin-spin interaction of the Ising type, defined on a square and a triangular lattice with OBC along the vertical direction, which can be embedded on a wide cylinder with a circumference much larger than its height. We probe the QPTs from the topological to the non-topological phase of the model occurring due to gradually increasing the perturbation strengths, and determine the QCPs using the fidelity susceptibility  of the ground state of the model, which exhibits a power-law divergence across the QPTs.  We verify the results using local magnetization of the 2D Ising model to which the perturbed Kitaev code can be mapped. Our investigation demonstrates an enhanced robustness of the topological phase of the Kitaev code against the local perturbations when OBC along the vertical direction is used, as compared to the case when the boundary condition along both directions are periodic. We also demonstrate a finite size odd-even dichotomy during the occurrence of the QPT depending on whether the circumference of the cylinder is even or odd, when the field perturbation is absent.   

We further consider a local entanglement witness operator constructed specifically to capture a lower bound to the localizable entanglement over a non-trivial vertical loop in the unperturbed Kitaev code, and demonstrate that the expectation value of the operator in the ground state of the perturbed Kitaev code exhibits a  signature of the QPT via a logarithmic divergence in its first derivative w.r.t. the local perturbation strength. This feature remains qualitatively unchanged in the case of the locally perturbed color code with a boundary condition similar to that of the Kitaev code, when the perturbation is in the form of local parallel magnetic field. The use of the witness operator makes our result potentially verifiable in experiments designed with various substrates including trapped ions~\cite{nigg2014,*linke2017} and superconducting qubits~\cite{Zhao2022,*gambetta2017}, where topological quantum codes can be realized.  

We conclude with a discussion on the challenges that lay ahead. Note that we have used exact diagonalization for obtaining the ground state of the locally perturbed topological codes as well as the models they map to, eg. the 2D Ising model in a transverse field~\cite{Rieger1999,Blote2002}, and the Baxter-Wu model in a transverse field~\cite{Capponi2014} with Ising interactions at the boundary. Therefore our study is still limited by the constraint of system size that is inherent in an exact diagonalization study. This is particularly prominent in the case of the locally perturbed color code on a tricolorable lattice with a system size growing faster compared to the Kitaev code, where scaling analysis with only the entanglement witness operator is performed by looking at the mapped model which provides an advantage in computation (see discussion in Sec.~\ref{subsec:kitaev_qpt_magnetization} around Eq.~(\ref{eq:effective_ising_square})). However,  obtaining enough data for performing a similar analysis with the fidelity susceptibility is not possible as it requires access to the actual ground state of the perturbed color code. Note that in the mapping of, for instance, the locally perturbed Kitaev code to the 2D Ising model, the designed local entanglement witness operator translates to an off-diagonal operator in the mapped model, and the determination of its expectation value using quantum Monte-Carlo technique itself is a difficult problem~\cite{Brower1998}, which needs to be addressed for probing larger systems. However, we point out here that this limitation does not hinder the thesis of this paper, and while one may indeed determine the QCPs and the critical exponents more precisely by employing a better numerical algorithm, our results are expected to hold qualitatively.          

\acknowledgments 

We acknowledge the use of \href{https://github.com/titaschanda/QIClib}{QIClib} -- a modern C++ library for general purpose quantum information processing and quantum computing. HKJ thanks the Prime Minister Research Fellowship (PMRF) program, Government
of India, for the financial support (PMRF ID: 3102904). The authors also thank Amit Jamadagni, Utkarsh Mishra, and Kedar Damle for useful discussions, and Anonymous Referee for insightful comments.  

\appendix 

\section{Scaling analysis}
\label{app:scaling}

Here, we briefly discuss the three-step procedure followed for determining the scaling exponents in this paper. The steps are as follows. 
\begin{enumerate}
    \item First, a non linear transformation of both abscissa and ordinate is performed as per the scaling formula (see Eqs.~(\ref{eq:fs_scaling_field}) and (\ref{eq:fs_scaling_ising})). For example, $\chi(g)\rightarrow (\chi(g_m)-\chi(g))/\chi(g)$ and $g\rightarrow D^{\nu_g}(g-g_m)$ in the case of fidelity susceptibility under field perturbation.
    \item Next, assuming a polynomial $\chi_g$ of order eight as the finite scaling function~\cite{Gu2008,*WingYu2009,Sandvik_2010},  we numerically perform the  least-square fitting of the re-scaled data to the finite scaling function $\chi(g)$.
    \item In the last step, we minimize the error in the least-square fitting by fine-tuning the critical exponents and critical perturbation strengths involved in the scaling function. We achieve this by numerical optimization tools.
\end{enumerate}  
For the last step, we exploit the connection between the perturbed toric code and the transverse-field Ising model to make a judicious choice of the range of values over which the search can be carried out. For example, we restrict the search space for the critical exponents within $\nu\in[0.622,1]$ and $\beta\in[0.125,0.314]$, resulting in a bounded optimization. This is motivated from the fact that the 1D transverse-field Ising model has the critical exponents $\nu=1,\beta=0.125$~\cite{Rieger1999,Sandvik_2010}, while the 2D transverse-field Ising model has the critical exponents $\nu\approx 0.622, \beta\approx 0.314$, as obtained using QMC simulations (see, for example,~\cite{Rieger1999}), and the model explored in this paper maps to a transverse-field Ising model on a quasi 1D lattice. The scaling analysis of all other quantities, namely the magnetization and the expectation value of the entanglement witness operator, are performed following the same procedure.

\begin{figure}
    \centering
    \includegraphics[width=1.0\linewidth]{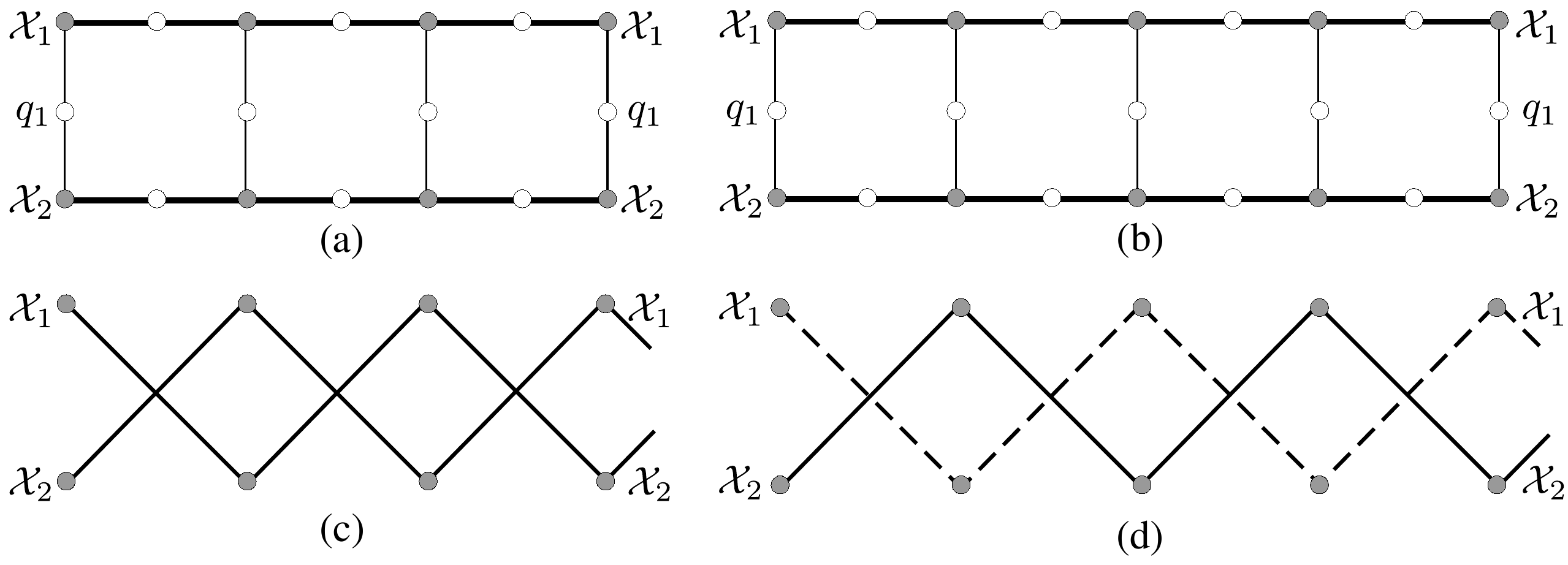}
    \caption{Kitaev's toric code on an $M=2$ square lattice with (a) $D=3$ and (b) $D=4$ can be mapped to respectively (c) one, and (b) two chains of qubits defined by the transverse-field Ising model when perturbed by nearest-neighbour longitudinal Ising interactions. The white (gray) circles denote physical qubits (vertices of the lattice) in the case of the Kitaev code where qubits reside on the edges. In the case of the transverse-field Ising chains, the lattice vertices host the qubits. In (d), the two different chains are denoted by solid and dashed lines. The labels $\mathcal{X}_1,\mathcal{X}_2$, and $q_1$ denotes the periodic boundary vertices and qubit respectively.}
    \label{fig:odd-even-dichotomy-mapping}
\end{figure}

\section{Finite size odd-even dichotomy} \label{app:odd_even_dichotomy}

For the quasi-1D lattice with $M=2$ and odd values of $D$, we perform rigorous numerical analysis of local magnetization of the mapped 1D quantum Ising model (see Eq.~(\ref{eq:effective_ising_square}) and corresponding discussions) using QMC techniques (see Fig.~\ref{fig:magnetization}), and reveal a QPT occurring at $\lambda_c^0=0.494$. Carrying out the the QMC simulation  for the square lattice in the cases of even $D$ becomes challenging due to the mapping of the lattice to two isolated Ising chains, and subsequent cancellation of magnetization (See Figure.~\ref{fig:odd-even-dichotomy-mapping}). However, in the limit $D\rightarrow\infty$, the existence of a QPT at a finite value of $\lambda$ is independent of whether the value of $D$ is even, or odd. To reveal the QPT for finite-size system in the case of even $D$, we explore the energy gap $\Delta \epsilon$ between the first excited state and the ground state,  FS, and the expectation value of the entanglement witness operator using the ED technique for both odd and even values of $D$. The findings are discussed below. 

\begin{figure*}
    \centering
    \includegraphics[width=1.0\linewidth]{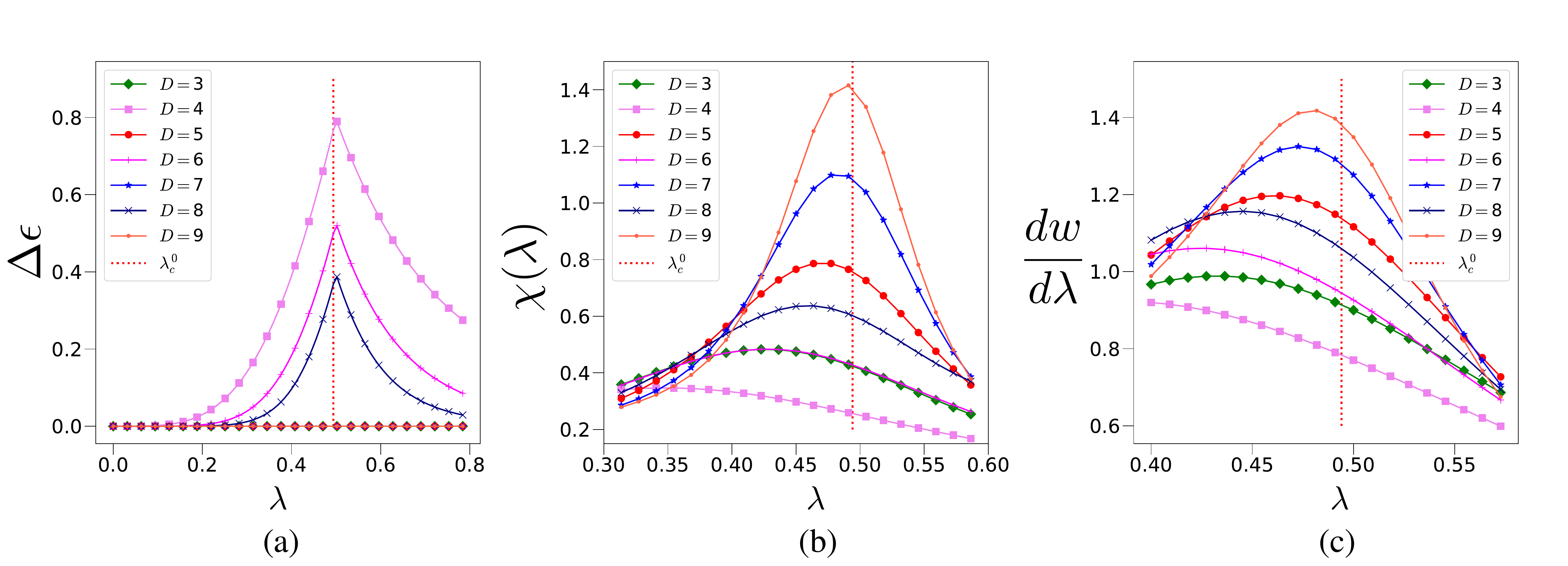}
    \caption{Variations of the (a) energy gap,  $\Delta\epsilon$, (b) fidelity susceptibility, $\chi(\lambda)$, and the (c) first-derivative of the expectation value of the witness operator, $w$, w.r.t. $\lambda$, as functions of $\lambda$ in the case of  an Ising-perturbed Kitaev code on an $M=2$ square lattice with variable $D$. The dashed vertical line indicates the critical perturbation strength $\lambda_c^0=0.494$ as obtained from the magnetization data using quantum Monte Carlo technique (see Table~\ref{tab:qpt}). All quantities plotted are dimensionless.}
    \label{fig:odd-even-dichotomy}
\end{figure*}

The variation of $\Delta\epsilon$ as a function of $\lambda$, depicted in Figure.~\ref{fig:odd-even-dichotomy}(a), indicates that $\Delta\epsilon=0$ for all odd values of $D$ for all finite $\lambda$ in the range $0\leq\lambda\leq\infty$. However, for small and even values of $D$, $\Delta\epsilon$ is finite for all finite $\lambda$ in the range $0\leq\lambda\leq \infty$, indicating a unique ground state (see also~\cite{Karimipour2013}). The non-analytic feature of $\Delta \epsilon$ with even $D$, caused due to an energy level-crossing,  occurs close to the critical point $\lambda_c^0=0.494$ predicted from the analysis involving odd values of $D$ and magnetization. However, with increasing $D$, $\Delta\epsilon\rightarrow 0$, indicating the independence in the behavior of $\Delta\epsilon$ of the choice of even, or odd values of $D$ in the limit $D\rightarrow\infty$.

We further investigate the FS, $\chi(\lambda)$, and the expectation value $w$ of the entanglement witness operator, the data for which are presented in Figs.~\ref{fig:odd-even-dichotomy}(b) and (c) respectively. In the case of the former, irrespective of whether $D$ is even or odd, a maximum is attained close to $\lambda_c^0=0.494$, although the approach of the maximum towards $\lambda_c^0$ depends on whether $D$ is even, or odd. For instance, the peak in $\chi(\lambda)$ becomes sharper and approaches $\lambda_c^0$ much faster in the case of odd $D$, as compared to the even values of $D$, which is clear from Fig.~\ref{fig:odd-even-dichotomy}(b), and is indicative of the fact that the even and odd values of $D$ may warrant for different scaling functions for $\chi(\lambda)$. While it is difficult to consider larger values of $D$ using ED technique, our finite-size analysis suggests $\lambda_c^0\approx 0.496$ when $D$ is odd, and $\lambda_c^0\approx0.508$ when $D$ is even, both of which are close to the value of  $\lambda_c^0$ obtained from odd values of $D$ in the limit $D\rightarrow\infty$. The qualitative features of the behavior of $dw/d\lambda$, as presented in Fig.~\ref{fig:odd-even-dichotomy}(c), is similar. It is worthwhile to mention that due to the nuance in the mapping of the Kitaev model to the transverse Ising model discussed above, the expectation values of the witness operator calculated from the transverse-field Ising model does not match with the same measured from the locally perturbed Kitaev model with even $D$, when $M=2$.

These results clearly indicate, for even values of $D$, the existence of  a quantum phase transition at a finite value of  $\lambda$ close to the value of $\lambda_c^0$ which is obtained using QMC techniques with odd values of $D$. Further, our analysis reveals subtle differences in the degeneracy of  the ground state, and in the approach of the system towards the limit $D\rightarrow\infty$, depending on whether odd, or even values of $D$ are considered.  We refer to this as the \emph{finite-size odd-even dichotomy}, which vanishes in the limit $D\rightarrow\infty$, as expected.

\section{Localizable entanglement}
\label{app:localizable_entanglement}

\begin{figure*}
    \centering
    \includegraphics[width=0.8\textwidth]{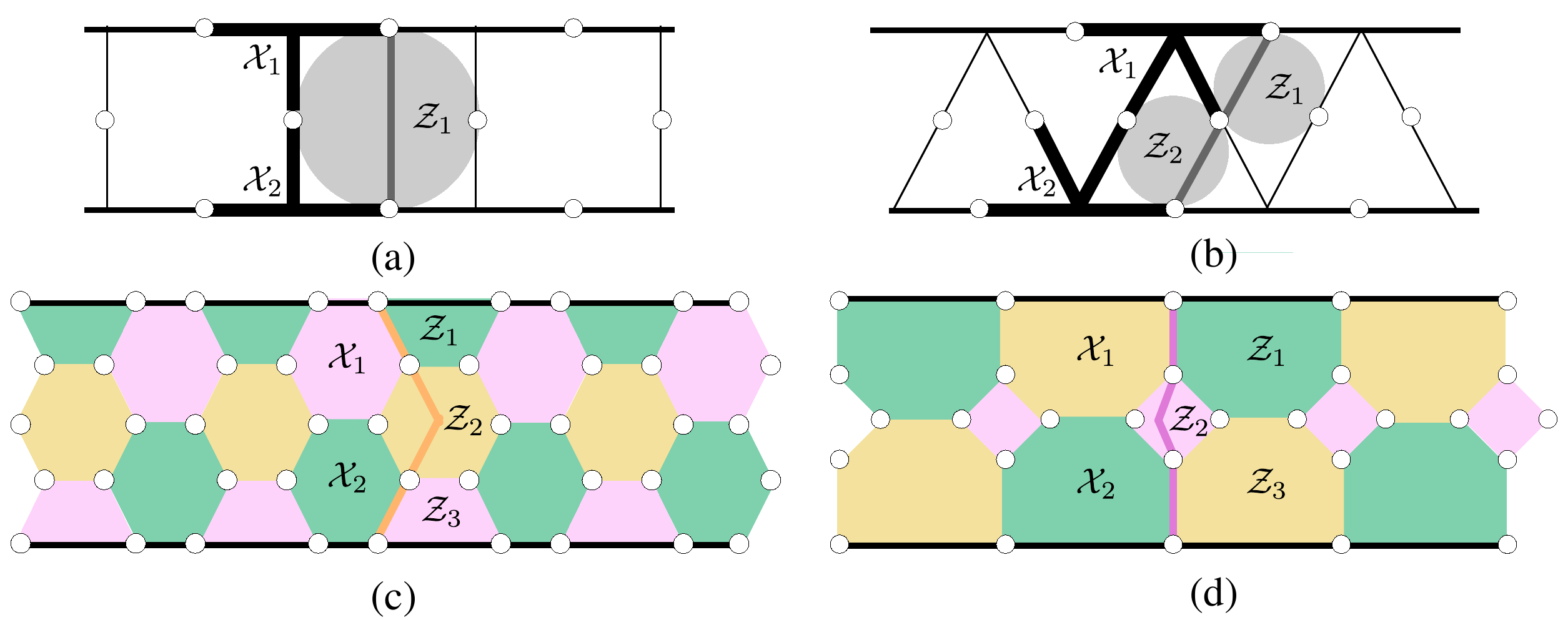}
    \caption{Construction of witness operators corresponding to (a) Kitaev code on square lattice with $M=2$, (b) Kitaev code on triangular lattice with $M=2$, (c) color code on honeycomb lattice with $M=3$, and (d) color code on square-octagonal lattice with $M=2$, as discussed in Sec.~\ref{subsec:kitaev_entanglement}. The vertical solid grey lines in the case of Kitaev codes, and the vertical solid lines in the case of the color codes indicate the non trivial loops along which entanglement is localized. For both the Kitaev and the color codes, contributing vertex (solid black lines for Kitaev code) and plaquette (gray circles for Kitaev code) operators are marked. The subset $\{\mathcal{S}_\alpha\}$ (see Eq.~(\ref{eq:witness}) and Appendix \ref{app:witness_construction}) in each of these cases will be as follows: (a) $\mathcal{S}_1=\mathcal{X}_1\mathcal{X}_2,\mathcal{S}_2=\mathcal{Z}_1$ with $n=2$, (b) $\mathcal{S}_1=\mathcal{X}_1\mathcal{X}_2,\mathcal{S}_2=\mathcal{Z}_1,\mathcal{S}_3=\mathcal{Z}_1\mathcal{Z}_2$ with $n=3$, (c) $\mathcal{S}_1=\mathcal{X}_1\mathcal{X}_2,\mathcal{S}_2=\mathcal{Z}_1,\mathcal{S}_3=\mathcal{Z}_1\mathcal{Z}_2,\mathcal{S}_4=\mathcal{Z}_1\mathcal{Z}_2\mathcal{Z}_3$ with $n=4$, and (d) $\mathcal{S}_1=\mathcal{X}_1\mathcal{X}_2,\mathcal{S}_2=\mathcal{Z}_1,\mathcal{S}_3=\mathcal{Z}_1\mathcal{Z}_2,\mathcal{S}_4=\mathcal{Z}_1\mathcal{Z}_2\mathcal{Z}_3$ with $n=4$.}
    \label{fig:witness_composition}
\end{figure*}

\textit{Localizable entanglement} (LE) is defined as the maximum average entanglement that can be localized over a subsystem $\Omega$ of a multi-qubit state $\rho$, by single qubit projective measurements on all qubits in the rest of the system $\overline{\Omega}$~\cite{popp2005,verstraete2004}:
\begin{equation}
    E^L=\max_{\{ \mathcal{M} \} } \sum_k p_k E(\rho_k),\label{eq:LE}
\end{equation}
where $p_k=\text{Tr}(\mathcal{M}_k \rho \mathcal{M}_k^\dagger)$ and $\rho_k=\text{Tr}_{\overline{\Omega}} (\mathcal{M}_k \rho \mathcal{M}_k^\dagger) / p_k$ are the probability and the corresponding post measured state in $\Omega$, respectively, for the measurement outcome $k$. The set $\{ \mathcal{M} \}$ denotes the set of all possible single qubit projective measurements in $\overline{\Omega}$. Here, $E$ is a bipartite or multipartite entanglement measure. A lower bound to $E^L$ can be obtained by judiciously fixing an appropriate measurement basis, say, an appropriate Pauli basis on every qubit in $\overline{\Omega}$, that provides a non-trivial value $E^\mathcal{P}$, where by the definition,
\begin{equation}
    E^\mathcal{P} \leq E^L.
\end{equation}
In the case of stabilizer states, the local unitary connection~\cite{van-den-nest2004} with graph states~\cite{Hein2005} can be exploited to arrive at a measurement setup that can lead to a non-trivial value of $E^\mathcal{P}$ for arbitrary choices of $\Omega$~\cite{HK2023}. In the case of the ground state of the Kitaev code (Eq.~(\ref{eq:kitaev})) one can choose a measurement setup such that $\rho_k$'s are all local unitary equivalent to Greenberger-Horne-Zeilinger (GHZ) states on the qubits in $\Omega$, where $\Omega$ is chosen to be a non-trivial loop of the code~\cite{HK2022,HK2023}. A similar result can also be obtained in the case of color code~\cite{HK2022}.

\section{Construction of local witness operators}
\label{app:witness_construction}

In this section, we briefly discuss the construction of local witness operators on a subsystem $\Omega$ of the topological code with $n \leq |L^x_v|$ number of qubits lying along the vertical non-trivial loop corresponding to the logical operator $\mathcal{L}^x_v$. A subset $\{\mathcal{S}_\alpha\}$ of all possible stabilizer operators can construct a local entanglement witness operator of the form Eq.~(\ref{eq:witness}), if it satisfies the following four conditions~\cite{Amaro2020}:
\begin{enumerate}
    \item The subset $\{\mathcal{S}_\alpha\}$ contains $n$ independent and commuting stabilizers.
    \item  The $n$ \textit{reduced Pauli operators} defined as $\{\mathcal{S}_\alpha^\Omega \}=\{Tr_{\overline{\Omega}}(\mathcal{S}_\alpha)\}$, where $\overline{\Omega}$ is the set of qubits outside $\Omega$, are independent, and mutually commute. 
    \item All the reduced single qubit Pauli operators, $\mathcal{S}_\alpha^{\overline{\Omega},i}$, on every qubit outside $\Omega$ mutually commute.
    \item The $n \times \genfrac(){0pt}{2}{n}{2} $ dimensional \textit{pseudoincidence matrix}$ (\mathcal{M})$ of the subsystem $\Omega$, defined as
    \begin{equation}
    \mathcal{M}_{i,j}=\begin{cases}
        $1,$  &\text{if}\hspace{0.5cm}  \{ \mathcal{S}^{\Omega,i}_\alpha,\mathcal{S}^{\Omega,i}_{\alpha^\prime}\}=0,\\
       $0,$  &\text{if}\hspace{0.5cm}  [ \mathcal{S}^{\Omega,i}_\alpha,\mathcal{S}^{\Omega,i}_{\alpha^\prime}]=0,
    		 \end{cases}    
    \end{equation}
    is a rank $n-1$ matrix. Here, $\mathcal{S}^{\Omega,i}_\alpha$ represents the single qubit reduced Pauli operator of the stabilizer $\mathcal{S}^{\Omega}_\alpha$ corresponding to qubit $i\in \Omega$ and $j\in[1,\genfrac(){0pt}{2}{n}{2}]$ is an index corresponding to a pair $\{\alpha,\alpha^\prime\}$ that runs through all possible stabilizer pairings within the set $\{\mathcal{S}_\alpha^\Omega\}$.
\end{enumerate}

We show that there exists a specific choice of such a subset $\{\mathcal{S}_\alpha\}$ for the Kitaev and the color codes such that all of the four conditions are satisfied. Consider a subset of stabilizer operators (or a certain recombination of them) such that only one of them is composed out of $x$ type stabilizers ($\mathcal{X}_V,\mathcal{X}_P$ for the Kitaev and the color codes, respectively), which we denote with $\mathcal{S}_1$. Every other element $\mathcal{S}_\alpha \forall \alpha >1$ is made out of only $z$ type stabilizers ($\mathcal{Z}_P$ for both the Kitaev and the color code). The contributing vertex and plaquette operators that construct the subset $\{\mathcal{S}_\alpha\}$ are chosen in such a way that the corresponding set of reduced Pauli operators, $\{\mathcal{S}_\alpha^\Omega \}$, takes the following form:
\begin{eqnarray}
    \mathcal{S}_1^\Omega&=&\otimes_{i=1}^{n}\sigma_i^x, \nonumber\\
    \mathcal{S}_\alpha^\Omega&=&\sigma_1^z\sigma_\alpha^z \forall \alpha\in\{2,3,\cdot\cdot\cdot n\}\label{eq:app_witness_form}.
\end{eqnarray}
See Fig.~\ref{fig:witness_composition} for an example of such a subset in the Kitaev and the color code with OBC along $v$ (see~\cite{HK2022} for a detailed discussion of the choice of $\{\mathcal{S}_\alpha\}$ in the case of PBC along both $h$ and $v$ directions). The subset $\{\mathcal{S}_\alpha\}$ (see Eq.~(\ref{eq:witness})) in each of the four cases depicted in Fig.~\ref{fig:witness_composition} are the following:
\begin{enumerate}
    \item For Kitaev code on square lattice with $n=2$ (Fig.~\ref{fig:witness_composition}(a)),  
    \begin{eqnarray}
    \mathcal{S}_1 &=& \mathcal{X}_1\mathcal{X}_2,\\
    \mathcal{S}_2 &=& \mathcal{Z}_1.
    \end{eqnarray} 

    \item For Kitaev code on triangular lattice with $n=3$ (Fig.~\ref{fig:witness_composition}(b)), 
    \begin{eqnarray}
    \mathcal{S}_1 &=& \mathcal{X}_1\mathcal{X}_2,\\
    \mathcal{S}_2 &=& \mathcal{Z}_1,\\
    \mathcal{S}_3 &=& \mathcal{Z}_1\mathcal{Z}_2    
    \end{eqnarray}
    \item In the cases of  color code on honeycomb and square-octagonal lattices (Figs.~\ref{fig:witness_composition}(c)-(d)) both having $n=4$,
    \begin{eqnarray}
    \mathcal{S}_1 &=& \mathcal{X}_1\mathcal{X}_2,\\
    \mathcal{S}_2 &=& \mathcal{Z}_1,\\
    \mathcal{S}_3 &=&\mathcal{Z}_1\mathcal{Z}_2,\\
    \mathcal{S}_4 &=&\mathcal{Z}_1\mathcal{Z}_2\mathcal{Z}_3.    
    \end{eqnarray}
\end{enumerate}

While constructing the pseudoincidence matrix $\mathcal{M}$, choosing a particular ordering for the pairing such that the first $n-1$ indices correspond to the pairings involving the $\mathcal{S}^\Omega_1$ component, i.e., $\{\{\alpha,\alpha^\prime\}\}=\{\{1,2\},\{1,3\},\cdot\cdot\cdot \{1,n\}\cdot\cdot\cdot$\} leads to $\mathcal{M}_{i,j}=0 \forall j \geq n$. Thus, the rank of $\mathcal{M}$ is equal to that of the $n\times (n-1)$ submatrix denoted by $\mathcal{M}^n$, involving only the first $n-1$ columns with nonzero entries. Due to the specific form of $\{\mathcal{S}_\alpha^\Omega\}$, it is straightforward to see that $\mathcal{M}^n_{1,j}=1\forall j$ and $\mathcal{M}^n_{i,j}=\delta_{i,j+1}\forall i>1$. As a specific example with $n=4$,
\begin{equation}
\mathcal{M}^4=
\begin{bmatrix}
    1 & 1 & 1 \\
    1 & 0 & 0 \\
    0 & 1 & 0 \\ 
    0 & 0 & 1 \\    
\end{bmatrix}.    
\end{equation}

Note that
\begin{enumerate}
    \item $\{\mathcal{S}_\alpha\}$ is composed of $n$ recombinations of plaquettes and vertices such that $[\mathcal{S}_\alpha,\mathcal{S}_{\alpha^\prime}]=0\forall \alpha,\alpha^{\prime}$. Hence, the first condition is satisfied.
    \item $\{\mathcal{S}^\Omega_\alpha\}$ contains $n$ independent operators (see Eq.~(\ref{eq:app_witness_form})) and $[\mathcal{S}^\Omega_\alpha,\mathcal{S}^\Omega_{\alpha^\prime}]=0\forall \alpha,\alpha^{\prime}$. Hence, the second condition is satisfied.
    \item The third condition is satisfied due to the freedom in choosing different plaquette recombinations such that all the reduced single qubit Pauli operators, $\mathcal{S}_\alpha^{\overline{\Omega},i}$, on every qubit outside $\Omega$ mutually commute.
    \item Finally, the fourth condition is satisfied, as we directly see rank$(\mathcal{M}^n)=n-1$. 
\end{enumerate} 
Thus, by the theorem in~\cite{Amaro2020}, all four necessary and sufficient conditions are satisfied, and $\{\mathcal{S}_\alpha\}$ constructs a local witness operator for the subsystem $\Omega$.

\section{Lower bound to localizable genuine multipartite entanglement with geometric measure}
\label{app:GM_witness_lb}
The set of \textit{reduced Pauli operators} $\{ \mathcal{S}_\alpha^\Omega\}$ defined in Appendix~\ref{app:witness_construction} form a complete set of stabilizer generators for the $n$-qubit GHZ state~\cite{greenberger1989} which we denote as $\ket{\text{GHZ}_n}$. Thus, the local witness operator defined by Eq.~(\ref{eq:witness}) and measured in the $N$-qubit state 
can provide a lower bound to LE over $\Omega$, which corresponds to a measurement strategy leading to post-measured states on $\Omega$ that are local unitary equivalent to $n$-qubit GHZ state $\ket{\text{GHZ}_n}$ ~\cite{HK2022,HK2023,amaro2018}. The local witness operator $W$ is equivalent to measuring a global witness operator of the form $1/2-\ket{\text{GHZ}_n}\bra{\text{GHZ}_n}$ in the subsystem $\Omega$. Given that $\langle W \rangle=w$, the witness-based lower bound to multipartite entanglement quantified in terms of geometric measure of entanglement~\cite{Barnum2001,*Wei2003}
is given by~\cite{guhne2007}
\begin{equation}
    E^w=\max_r f(r,w),
\end{equation}
with 
\begin{equation}
    f(r,w) = \begin{cases}
        rw+\frac{1-\sqrt{1+r^2}}{2}, \text{  if  } r<0,\\
        \frac{r}{2}, \text{  if  } r>0,
    \end{cases}
\end{equation}
where we have used the fact that the value of the geometric measure of entanglement for the GHZ state is $1/2$. Maximizing $f$ w.r.t. $r$ leads to Eq.~(\ref{eq:lb_gme}) at $r=-2/\sqrt{\omega^{-2}-4}$.

\nocite{apsrev41Control}
\bibliography{ref}

\end{document}